\documentclass[aps,prl,reprint,groupedaddress]{revtex4-1}

\usepackage{graphicx,amsmath,color,amsfonts,amsmath,bm}

\usepackage[colorlinks=true]{hyperref}  
\hypersetup{
    bookmarks=true,         
    unicode=false,          
    pdftoolbar=true,        
    pdfmenubar=true,        
    pdffitwindow=false,     
    pdfstartview={FitH},    
    pdftitle={},    
    pdfauthor={},     
    pdfsubject={},   
    pdfcreator={},   
    pdfproducer={}, 
    pdfkeywords={} {} {}, 
    pdfnewwindow=true,      
    colorlinks=true,       
    linkcolor=blue, 
    citecolor=blue,        
    filecolor=magenta,      
    urlcolor=blue,           
        breaklinks=true
}

\newcommand{\be}{\begin{equation}}
\newcommand{\ee}{\end{equation}}
\newcommand{\bea}{\begin{eqnarray}}
\newcommand{\eea}{\end{eqnarray}}

\begin{document}

\title{Transverse instability and universal decay of spin spiral order in the Heisenberg model}

\author{Joaquin F. Rodriguez-Nieva$^{1}$, Alexander Schuckert$^{2,3}$, Dries Sels$^{4,5,6}$, Michael Knap$^{2,3}$, Eugene Demler$^6$}

\affiliation{$^1$Department of Physics, Stanford University, Stanford, CA 94305, USA}

\affiliation{$^2$Department of Physics and Institute for Advanced Study,Technical University of Munich, 85748 Garching, Germany}

\affiliation{$^3$Munich Center for Quantum Science and Technology (MCQST), Schellingstr. 4, D-80799 M{\"u}nchen}

\affiliation{$^4$Department of Physics, New York University, New York, NY, USA}

\affiliation{$^5$Center for Computational Quantum Physics, Flatiron Institute, New York, NY, USA}

\affiliation{$^6$Department of Physics, Harvard University, Cambridge, MA 02138, USA}

\date{\today}

\begin{abstract}

  We analyze the stability of spin spiral states in the two-dimensional Heisenberg model. Our analysis reveals that the SU(2) symmetric point hosts a dynamic instability that is enabled by the existence of energetically favorable transverse deformations---both in real and spin space---of the spiral order. The instability is universal in the sense that it applies to systems with any spin number, spiral wavevector, and spiral amplitude. Unlike the Landau or modulational instabilities which require impurities or periodic potential modulation of an optical lattice, quantum fluctuations alone are sufficient to trigger the transverse instability. We analytically find the most unstable mode and its growth rate, and compare our analysis with phase space methods. By adding an easy plane exchange coupling that reduces the Hamiltonian symmetry from SU(2) to U(1), the stability boundary is shown to continuously interpolate between the modulational instability and the transverse instability. This suggests that the transverse instability is an important mechanism that hinders the formation of a spin superfluid, even in the presence of strong exchange anisotropy. 
\end{abstract}



\maketitle

Characterizing the mechanisms responsible for the breakdown of phase coherence in quantum systems is a fundamental problem with broad implications in quantum science and technology. The interplay between kinetic effects, interactions, and disorder gives rise to a wide range of phase relaxation mechanisms. In the simplest scenario, the phase coherence in a superfluid is subject to the Landau criterion\cite{1941landaucriterion} which defines an upper limit for the superfluid velocity: when the superfluid moves faster than the sound velocity, a spatially localized defect can trigger a superfluid instability that globally destroys phase coherence\cite{2001sciencelandauinstability}. In the case of a Bose-Einstein condensate (BEC) in an optical lattice with spacing $a$, the characteristic lattice modulation $q_{\rm l}=\pi/2a$ sets another limit for the superfluid wavevector above which a modulational instability occurs\cite{2001niukinematic,2004prlmodulationalinstability}. Such instability can be enhanced in the presence of strong interactions\cite{2004polkovnikovmott,2005altmanmott}. Rich physics and diverse mechanisms that destroy---and sometimes stabilize---the phase coherence have been discussed in the context of counterflowing superfluids\cite{2005prlgardiner}, multicomponent\cite{2010prltsubota} and spinor BECs\cite{2008instabilityxy,2009spinorbec,2012pratsubota1}, superconductors\cite{1967scdecay,1970scdecay,1998so5scdecay,2010scdecayreview}, in the presence of extended disorder\cite{2007becdisorder,2008becdisorder,2010becdisorder,2019ueda}, dipolar interactions\cite{2000dipolarbec,2002dipolarbec,2003dipolarbec}, and driving\cite{2001parametricpumping,2018magnoninstabilities,2018heatingdrivenbec,2015bukovfloquet,2019instabilitydrivenbec}.

\begin{figure}[b]
\includegraphics[scale = 1.0]{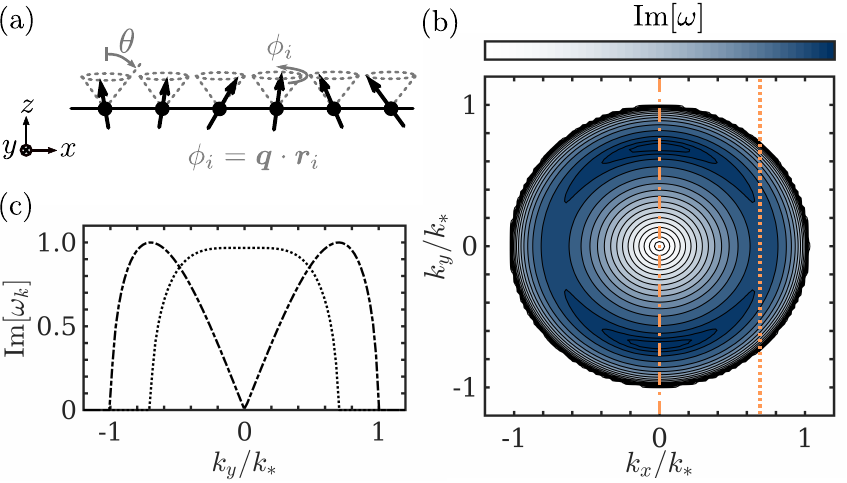}
\caption{(a) Schematics of a spin spiral parametrized by a wavevector ${\bm q}$ and angular amplitude $\theta$. (b) Imaginary part of the frequency of Bogoliubov modes. The wavevectors $k_{x,y}$ are relative to $k_* = q\sin\theta$, and ${\bm q}$, which is assumed to be pointing in the $x$-direction ($k_* = q\sin\theta$). The fastest growing modes are   transverse to $\bm q$, with $k_y \approx k_*/\sqrt{2}$. (c) Slice of panel (b) plotted at the linecuts $k_x/k_* = 0$ (dotted-dashed line) and $k_x/k_* = 0.75$ (dotted line) and normalized with $1/\tau_*$ in Eq.(\ref{eq:scaling}). Parameters used: $\theta = \pi/4$, $q_x a= 0.5$, $q_y=0$.}
\label{fig:schematics}
\end{figure}

Here we inquire about the fate of a spin spiral state in the two-dimensional Heisenberg model, see Fig.\,\ref{fig:schematics}(a). Understanding the stability and dynamics of such states is of relevance in many important scenarios. The non-equilibrium dynamics of spin spirals has recently been in the spotlight of several cold atom experiments\cite{2014spiralscience,2014spiralexp,2015sciencespiral}. By tuning the wavevector ${\bm q}$ and angular amplitude $\theta$ of the spin spiral, we can tune the energy and magnetization of the system and trigger interesting far-from-equilibrium phenomena, such as quantum turbulence \cite{2009tsubotareview,2020turbulence}, prethermalization\cite{2015PRX-babadi}, universal self-similar relaxation\cite{2015asier,2015bergesreview}, and anomalous transport\cite{2020spiralketterle}. In addition, the stability of spin superfluids in ferromagnetic materials, promising for dissipationless spintronic applications\cite{spinsuperfluidreview}, hinges on the stability of long-range coherence of a spin spiral. Our results will uncover (in a sense that we specify below) the fragility of the spin superfluid state, justifying the scarce experimental evidence\cite{2016supercurrentyig,2016sonincomment,2016hillebrandsreply} in contrast to all the theoretical investigations (for example, \cite{2014yaroslavsuperfluid,2014lossspinsuperfluidity,2016yigsuperfluidity,2017spinsuperfluiditysonin}). 

Our analysis reveals that the SU(2) symmetry of the Hamiltonian gives rise to a dynamic instability with different characteristics from previously-studied instabilities. In particular, the instability (i) is enabled by gapless symmetry-allowed deformations of the order parameter rather than kinematic effects, (ii) is triggered by quantum fluctuations without the need for defects, disorder or a lattice, and (iii) is universal in the sense that it affects systems with arbitrary spin number $S$, spiral wavevector, and spiral amplitude. The main physics can be understood by noticing that the SU(2) symmetry relaxes the topological constraint that protects the U(1) phase in superfluids\cite{spinsuperfluidreview}: while in usual superfluids the thermally-activated creation of vortex-antivortex pairs or large kinematic fluctuations destroy coherence, the SU(2) symmetry alone furnishes additional `directions' (or rotation generators) in which the phase coherence can be destroyed. As indicated in Fig.\,\ref{fig:schematics}(b), the instability evolves by unwinding the spiral via growth of modes in a ring around the wavevector $\bm q$. Assuming ${\bm q} = (q_x,0)$, the fastest growing mode has transverse wavevector $k_\perp \approx k_*/\sqrt{2}, $ and grows with a rate $1/\tau_*$, with 
\be
k_* =  |\bm q|\sin\theta, \quad \frac{1}{\tau_*} = JS\sin^2\theta[1-\cos({q}_xa)],
\label{eq:scaling}
\ee
and $J$ the exchange coupling. In addition, numerical simulations show that the constraint $\hat{S}_i^2 = S(S+1)$ of each spin regulates the instability growth, which peaks in a time $t\approx 4\tau_*$ (largely independent of ${\bm q}$, $\theta$, and $S$). 

We analytically discuss dynamics in the SU(2) symmetric Heisenberg model for large $S$, but our conclusion are far more general. In particular, below we numerically show that the imprint of the ring of unstable modes survives even in the $S=1/2$ limit for sufficiently small wavevectors. In addition, we show that the effect of the instability pervades away from the SU(2) symmetric point. Indeed, in the presence of anisotropic exchange that reduces the Hamiltonian symmetry to U(1), we observe a strong reduction of the critical wavevector for modulation instabilities ({\it i.e.}, $|\bm q_c| = \pi/2a$) for a wide range of values of the exchange anisotropy.

{\bf Microscopic model}---We consider the two-dimensional Heisenberg model on a square lattice with exchange anisotropy: 
\be
\hat{H} = -\sum_{\langle i,j \rangle} J\left(\hat{S}_i^x\hat{S}_j^x+\hat{S}_i^y\hat{S}_j^y\right)+ J_z \hat{S}_{i}^z\hat{S}_j^z,
\label{eq:hamiltonian}
\ee
where $\langle i , j \rangle$ denotes summation over nearest neighbours. Each site contains a spin $S$ degree of freedom and periodic boundary conditions in each spatial direction are assumed. The sign of $J$ does not affect the unitary evolution due to time reversal symmetry of $\hat{H}$. 
Our analysis is not affected by a Zeeman field, which is present in many relevant experiments: although a Zeeman field breaks the SU(2) symmetry of the Hamiltonian, its effect on dynamics can be removed by using a rotating frame. The initial condition is a spin spiral 
\be
\langle \hat{S}_i^\pm\rangle = S\sin\theta e^{\pm i{\bm q}\cdot{\bm r}_i},\quad \langle \hat{S}_i^z \rangle = S\cos\theta, 
\label{eq:initial}
\ee
with $\hat{S}_i^\pm = \hat{S}_i^x\pm i \hat{S}_i^y$. 

{\bf Bogoliubov analysis}---The equations of motion of the spin operators are given by $\partial_t\hat{\bm S}_i = J\sum_{j\in {\cal N}_i}\hat{\bm S}_i\times (\hat{\bm S}_j+\epsilon\hat{S}_j^z{\bm z})$, with $\epsilon=(J_z-J)/J$,  ${\cal N}_i$ the nearest neighbors of site $i$, and $\bm z$ a unit vector. We first analyze dynamics in the large $S$ limit using the approximation $\langle \hat{S}_i^\alpha \hat{S}_{j}^\beta\rangle \approx \langle \hat{S}_i^\alpha \rangle\langle \hat{S}_{j}^\beta\rangle $, which gives rise to the equations of motion 
\be
\begin{array}{l}
\displaystyle  \dot{S}_i^\pm = \mp i J\sum_{j\in{\cal N}_i}\left[(1+\epsilon)S_i^\pm S_j^z -S_j^\pm S_i^z\right],\\ \displaystyle\dot{S}_i^z = \frac{iJ}{2}\sum_{j\in{\cal N}_i}\left[ S_i^+S_j^--S_i^-S_j^+\right],
  \end{array}
\label{eq:eom}
\ee
with $\langle \hat{S}_i^\alpha\rangle = S_i^\alpha$. Hereafter, energy and inverse time are expressed in units of $JS$ and wavevectors in units of $1/a$. Using the initial conditions in Eq.\,(\ref{eq:initial}), it can be shown that the solution $\bar{S}_i^\pm(t) = S\sin\theta e^{\pm i ({\bm q}\cdot{\bm r}_i + \mu t)}$, $\bar{S}_i^z = S\cos\theta$, is a steady-state solution of Eq.(\ref{eq:eom}), with oscillation frequency $\mu = 2\cos\theta[(1+\epsilon)2 - \cos q_x-\cos q_y]$. Therefore, one needs to incorporate quantum fluctuations to obtain non-trivial dynamics.  

We proceed to analyze the stability of the spiral in the isotropic exchange case, $\epsilon=0$. We parametrize fluctuations on top of the steady-state solution using the $xy$ components of magnetization, $S_i^\pm = \bar{S}_i^\pm + \delta S_i^\pm$; this implies that our parametrization is singular at $\theta = \pi/2$, but taking the limit $\theta \rightarrow \pi/2$ at the end still yields the correct result (a parametrization in polar coordinates that is non-singular at $\theta=\pi/2$, but more cumbersome, is discussed in the Supplement). Going into momentum space and expressing modes relative to the wavevector and frequency of the spiral, $\delta S_i^\pm = e^{\pm i({\bm q}\cdot {\bm r}_i+\mu t)}\sum_{\bm k}e^{(i{\bm k}\cdot {\bm r}_i+\omega_{\bm k}t)}\delta S_{\bm k \pm \bm q}^\pm$, the linearized equations of motion are (see Supplement)  
\be
\left(\begin{array}{cc} \omega_{\bm k}+\varepsilon_{\bm q + \bm k}+\frac{\Delta_{\bm k}}{2}-\mu & \frac{\Delta_{\bm k}}{2} \\ -\frac{\Delta_{\bm k}}{2} & \omega_{\bm k} - \varepsilon_{\bm q - \bm k} -\frac{\Delta_{\bm k}}{2}+\mu\end{array}\right) \delta{\bm S}= 0.
\label{eq:bogoliubov}
\ee
Here $\delta{\bm S} = (\delta S_{\bm q + \bm k}^+ , \delta S_{\bm k - \bm q}^-)^{\rm t}$, and $\varepsilon_{\bm p} $, $\Delta_{\bm k}$ are 
\be
  \varepsilon_{\bm q\pm\bm k} = \cos\theta \left(\gamma_0-\gamma_{\bm q\pm\bm k}\right),\quad
 \Delta_{\bm k} = \sin\theta\tan\theta (\gamma_{\bm q} - \gamma_{\bm k}).
\label{eq:parameters}
\ee
where we defined $\gamma_{\bm k} = 2(\cos k_x + \cos k_y)$. Note that the Bogoliubov analysis can be easily generalized to next nearest neighbor interactions by modifying the definition of $\gamma_{\bm k} $ accordingly. From Eq.(\ref{eq:parameters}), we note that the value of $\mu $ is $\mu = \varepsilon_{\bm q}$. The frequencies of the Bogoliubov modes are 
\be
\omega_{\bm k} = \frac{\varepsilon_{\bm q + \bm k}-\varepsilon_{\bm q - \bm k}}{2}\pm\frac{1}{2}\sqrt{\Delta\varepsilon(\Delta\varepsilon+2\Delta_{\bm k})},
\label{eq:frequency}
\ee
where $\Delta\varepsilon=\varepsilon_{\bm q + \bm k} + \varepsilon_{\bm q - \bm k} - 2\varepsilon_{\bm q}$ can be interpreted as the kinetic energy cost of unbinding two quasiparticles from mode $\bm q$. For large spiral wavevectors, $q_x,q_y>\pi/2$, $\Delta\varepsilon$ can be negative because of the negative mass of bare particles and gives rise to the previously studied  modulational instability\cite{2001niukinematic}. For $q_x,q_y<\pi/2$, however, $\Delta\varepsilon$ is strictly positive and the condition for the mode $S_{\bm q+\bm k}^+$ to be unstable ({\it i.e.}, $\omega_{\bm k}'' = {\rm Im}[\omega_{\bm k}]\neq 0$) is given by
\be
\varepsilon_{\bm q+\bm k}+\varepsilon_{\bm q-\bm k}-2\varepsilon_{\bm q} + 2\Delta_{\bm k}<0.
\label{eq:balance}
\ee

It is instructive to analyze the condition (\ref{eq:balance}) in the limit of small $\bm q$ and $\theta$, and contrast it with the usual Landau instability. In this case, Eq.(\ref{eq:balance}) is an energy balance equation resulting from unbinding two magnons of energy $\varepsilon_{\bm q \pm \bm p} \approx JSa^2(\bm p \pm \bm q)^2$ and with a {\it momentum-dependent} pairing energy $\Delta_{\bm k} \approx -JSa^2 \sin^2\theta({\bm q}^{2}-{\bm k}^2)$. Importantly, $\Delta_{\bm k}$ is attractive in a ring of radius $|\bm k|\lesssim|\bm q|$. Attractive magnon-magnon interactions are known to give rise to magnon bound states in 1D\cite{mattisbook} and its momentum dependence has been shown to result in unusual quasiparticle relaxation\cite{2019saraswat} and hydrodynamic behavior\cite{2017breakinggalilean,2018hydrodynamicsoundmodes}.  Equation (\ref{eq:balance}) dictates that, independently of $S$, growth of modes with small wavevectors $\bm k$ relative to ${\bm q}$ is energetically favorable (a large value of $\bm k$, on the other hand, is penalized by a large kinetic energy cost, $\varepsilon_{\bm q+\bm k}+\varepsilon_{\bm q-\bm k}-2\varepsilon_{\bm q}\propto {\bm k}^2$). Condition (\ref{eq:balance}) needs to be contrasted with the superfluid stability condition where $\Delta_{\bm k} = gn>0$ is momentum independent and repulsive, thus ensuring stability of the superfluid in the long wavelength limit ($g$: local interaction, $n$: density). 

More generally, Eq.(\ref{eq:balance}) gives rise to unstable modes for any value of ${\bm q}$ and $\theta$. To analytically find the most unstable mode when ${\bm q} = (q_x,0)$, we maximize $\omega_{\bm k}''$ under the constraint $k_x = 0$ [note that the fastest growing mode in Fig.\,\ref{fig:schematics}(a) is transverse to $\bm q$]. In this case, we obtain 
\be
   \omega_{\bm k}'' = 2\sqrt{(1-\cos k_y)\left[(1-\cos k_y)-\sin^2\theta\left(1-\cos q_x\right)\right]}.
   \label{eq:imagomega}
\ee
From this, we see that the maximum of $\omega_{\bm k}''$ occurs at $k_y = K_y$, with $K_y$ satisfying $1-\cos K_{y} = \sin^2\theta(1-\cos q_x)/2$, and such mode grows with a rate ${\rm max}\left(\omega_{\bm k}''\right) = \frac{1}{\tau_*}$ in Eq.(\ref{eq:scaling}). Equation (\ref{eq:imagomega}) also defines the volume in phase space of unstable modes, which is bounded by the wavevector $k_*$ satisfying the condition $1-\cos k_* = \sin^2\theta[1-\cos q_x]$. In the limit of small $q_x$, we obtain $K_{y} \approx k_*/\sqrt{2}$, with $k_* $ defined in Eq.(\ref{eq:scaling}).

\begin{figure}
  \includegraphics[scale = 1.0]{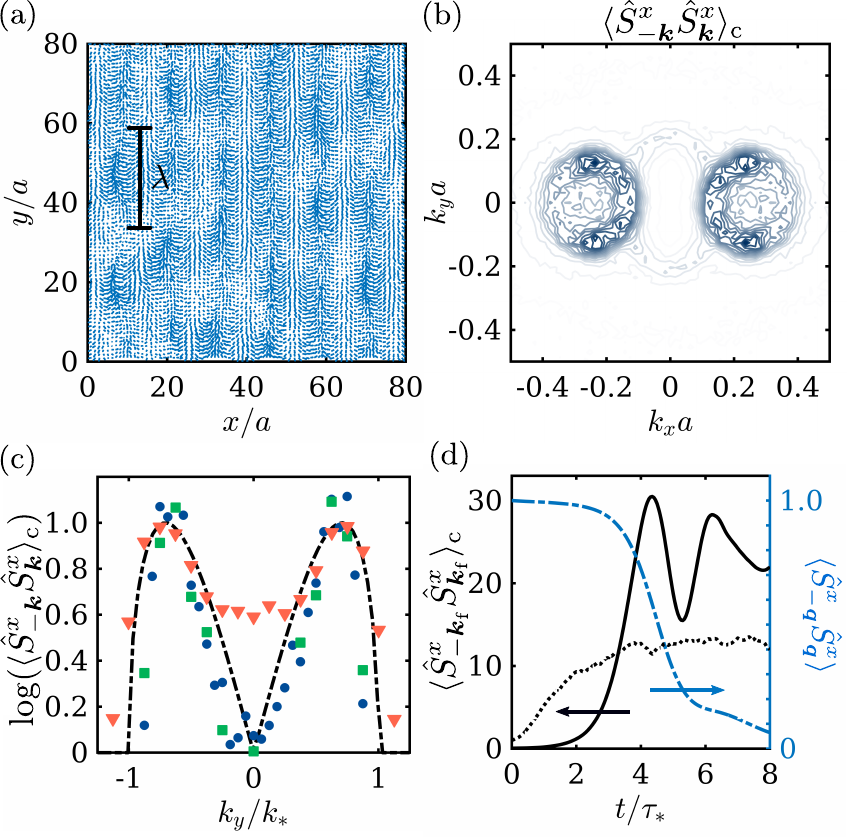}
  \vspace{-3mm}
  \caption{Real space snapshot of a single TWA realization at $t/\tau_* = 3.5$, see Eq.(\ref{eq:scaling}).  Shown are snapshots of the spins $ \hat{\bm S}_i$ projected on the $xy$ plane. Indicated with a bar is the wavelength in the $y$ direction of the fastest growing mode. (b) Contour plot showing the connected correlation $\langle \hat{S}_{-\bm k}^x\hat{S}_{\bm k}^x\rangle_{\rm c}$ corresponding to panel (a). Consistent with the Bogoliubov analysis, the plot exhibits a ring of unstable modes around ${\bm q}$ with size $k_*\approx q_x\sin\theta$ and a maximal amplitude transverse to $\bm q$. Parameters used in panel (a,b): $\theta = \pi/4$, $q_x = 0.25$, $q_y=0$, $S=10$ averaged over 50 realizations. (c) Spatial-temporal scaling of $\langle \hat{S}_{-\bm k}^x\hat{S}_{\bm k}^x \rangle_{\rm c}$. Shown is the log of $\langle \hat{S}_{-\bm k}^x\hat{S}_{\bm k}^x \rangle_{\rm c}$ for fixed $k_x=q_x$ and $t/\tau_*=3.5$, and for different initial conditions: $(q_x,\theta) = (0.25,\pi/2)$ (circles), $(0.5,\pi/2)$ (squares) and $(0.25,\pi/4)$ (triangles) and $S=10$. The dotted-dashed line is the Bogoliubov ${\rm Im}[\omega]$ in Eq.(\ref{eq:imagomega}) as a guide to the eye. (d) Growth of the most unstable mode, ${\bm k}_{\rm f} = (q_x,K_y)$, showing saturation and subsequent oscillations for $S=10$ (solid line) and $S=1/2$ (dotted lines). Also shown is the depletion of the spin spiral (dashed-dotted line).
  }
\label{fig:relaxation}
\end{figure}

\begin{figure}[t]
  \includegraphics[scale = 1.0]{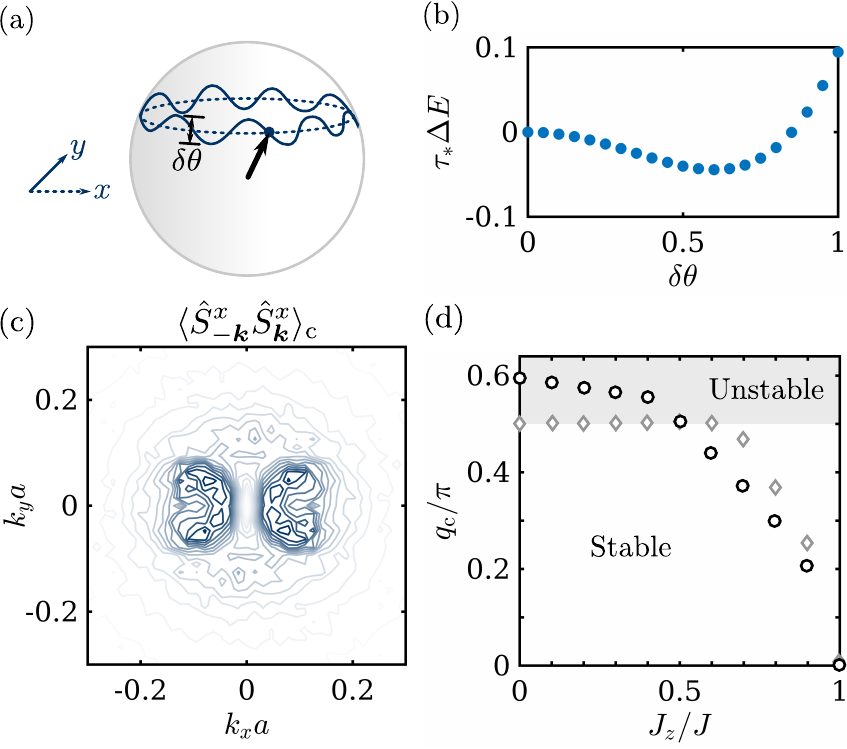}
  \vspace{-3mm}
  \caption{(a) Schematics of the most favorable deformation of the spin spiral order and (b) the corresponding energy shift with respect to the spiral state. Shown in (a) is the trajectory of the magnetization vector by moving on the lattice in the $x$ (dotted line) and $y$ (solid line) directions, and $\delta\theta$ is the amplitude of deformation, see definition in main text.
    (c) Contour plot showing the connected component of $\langle \hat{S}_{-\bm k}^x\hat{S}_{\bm k}^x\rangle$ for $S=1/2$, $q_xa = 0.12$, and $\theta = \pi/2$. (d) Stability boundary showing the critical momentum $q_{\rm c}$ as a function of exchange anisotropy for a spiral amplitude of $\theta = \pi/4$ (black circles), $\theta=0.1$ (gray diamonds), and $S\rightarrow\infty$. The shaded area indicates the parameter space region of the modulational instability. }
\label{fig:boundary}
\end{figure}

{\bf Phase space methods}---To complement the Bogoliubov analysis, we compute real time dynamics of the spiral decay by incorporating quantum fluctuations using the Truncated Wigner Approximation\cite{2010polkovnikovreview}. Defining $\langle \hat{\bm S}_i^\perp \rangle$ as the transverse magnetization of the initial condition (\ref{eq:initial}), we assume Gaussian fluctuations of $\hat{\bm S}_i^\perp$ given by $\langle\hat{\bm S}_i^\perp\rangle = 0$ and $\langle\hat{\bm S}_i^\perp\cdot\hat{\bm S}_i^\perp\rangle = S$.

Figure \ref{fig:relaxation}(a) shows a single realization of TWA noise for a spin spiral with parameters $\theta = \pi/4$ and $q_x = 0.5$ (same parameters as in Fig.\ref{fig:schematics}). Independently of the spin number $S$, we consistently observe growth of unstable modes that lead to a disordered state. Analysis of the connected correlation $\langle \hat{S}_{-\bm k}^x\hat{S}_{\bm k}^x\rangle_{\rm c} = \langle \hat{S}_{-\bm k}^x\hat{S}_{\bm k}^x\rangle - \langle \hat{S}_{-\bm k}^x\rangle\langle\hat{S}_{\bm k}^x\rangle$ [shown in Fig.\ref{fig:relaxation}(b)] reveals that the spiral state is primarily decaying into modes located in a ring around the wavevector ${\bm q}$, preferentially in the direction perpendicular to ${\bm q}$, thus confirming the Bogoliubov analysis above.

In addition, Fig.\ref{fig:relaxation}(c) shows the scaling of fluctuations for wavectors $\bm k = (q_x,k_y)$ and various initial conditions at the rescaled time $t/\tau_* = 3.5$. Given that we expect unstable modes to grow as $S_{\bm k}^+(t) \approx S_{\bm k}^+(0)e^{t/\tau_{\bm k}}$, the $y$ axis is plotted in log scale and the correlation $\langle \hat{S}_{-\bm k}^x \hat{S}_{\bm k}^x\rangle_{\rm c}$ is normalized with the maximum value as a function of $k_y$ for each initial condition. We observe excellent agreement with the Bogoliubov analysis  for all 
$\bm q$ and $\theta$. 

{\bf Instability growth and self-regularization}---Going beyond the linear stability analysis, we inquire about the intermediate timescale dynamics of instability growth. Figure\,\ref{fig:relaxation}(d) shows the decay of the spin spiral and multiple stages in the evolution of the most unstable mode: (i) initial growth compatible with the Bogoliubov analysis above, (ii) saturation, (iii) coherent oscillations prior to equilibration. Unlike usual instabilities in BEC where unstable modes grow exponentially for long times, the local constraint $\hat{S}_i^2 = S(S+1)$ and the conservation of total magnetization regulates the growth of the transverse spin modulation at relatively short times, analogously to Refs.[\onlinecite{2015PRX-babadi,2003prlberges,2008berges}]. We observe that saturation occurs at $t \approx 4\tau_*$, irrespective of the value of $S$, $\bm q$ and $\theta$ (see Supplement).  

The existence of unstable modes in the linearized analysis and the small oscillations in Fig.\ref{fig:relaxation}(d) are linked to the existence of smooth, symmetry-allowed deformations of the spin spiral order with a valley-shaped potential. Using the insights gained from the Bogoliubov analysis, we propose a simple Ansatz for a transverse spin texture given by $S_i^\pm = S \sin\theta_i e^{\pm i {\bm q}\cdot{\bm r}_i}$ and $S_i^z =S \cos\theta_i$, with $\theta_i = \bar{\theta}+2\delta\theta\cos(K_yy_i)$ and $K_y$ defined below Eq.(\ref{eq:scaling}) [see Fig.\ref{fig:boundary}(a)]. The value of $\delta\theta$ controls the amplitude of transverse spin deformations around $\bar{\theta}$ and is modulated by the transverse wavevector $K_y$. This ansatz trivially satisfies $\sum_i S_i^\pm = 0$ for all values of $\bar\theta$ and $\delta\theta$, and the condition $\frac{1}{N}\sum_i S_i^z = S\cos\theta$ defines a constraint that links $\bar\theta$ and $\delta\theta$. Because we recover Eq.(\ref{eq:initial}) when $\bar\theta = \theta$ and $\delta\theta = 0$, our Ansatz is smoothly connected to the original spiral and preserves its total magnetization. Figure\,\ref{fig:boundary}(b) shows that increasing the transverse modulation $\delta\theta$ reduces the energy of the spin spiral. In addition, the observed oscillations in Fig.\ref{fig:relaxation}(d) can be interpreted as amplitude oscillations on a valley-shaped potential. The same argument can be applied to any value of ${\bm k}$ that satisfies the instability condition (\ref{eq:balance}), but the valley is deepest for $\bm k = (q_x,K_y)$. 

{\bf Crossover to the quantum regime}---
The stability analysis above relies on a $1/S$ expansion of the equations of motion, opening the question on its validity in the experimentally relevant $S=1/2$ case. The competition between quantumness in the $S\rightarrow 1/2$ limit and classicality in the $\bm q\rightarrow 0$ limit suggests that a smeared, but still observable, ring of unstable modes is obtained for finite but small $\bm q$ and $S=1/2$. Indeed, our numerics reveal that strong quantum fluctuations supress the exponential growth of unstable modes and smear out coherent oscillations in the two-point correlation function [see Fig.\ref{fig:relaxation}(d)], but the latter still exhibits an imprint of the ring of unstable modes, see Fig.\ref{fig:boundary}(c). Remarkably, we also find that our simple semiclassical picture essentially survives in the one-dimensional Heisenberg model despite integrability and reduced dimensionality, as shown with Matrix Product States in the Supplement (in this case, the most unstable modes are necessarily collinear with $q_x$). 

{\bf Crossover to modulational instability}---
To study the crossover between the transverse instability in the Heisenberg model ($J_z = J$) to the modulational instability that characterizes a superfluid on a lattice, we extend the Bogoliubov analysis for values of $J_z < J$ (see details in the Supplement). Tuning $J_z$ can be realized experimentally using Feshbach resonances, dipolar interactions or lattice shacking\cite{2008amosuperexchange,2013amodipolarexchange,2013amoferromagnet,2016hungexchange,2019amospin}. The anisotropic exchange energetically penalizes the transverse deformation of the spin spiral order. In the language of the stability condition in Eq.(\ref{eq:balance}), the pairing $\Delta_{\bm k}$ becomes repulsive, $\Delta_{\bm k}\approx (J-J_z)$. While breaking the SU(2) symmetry has a stabilizing effect on the spin spiral state, there is still a strong reduction of the critical wavevector far from the SU(2) symmetric point, as shown in Fig.\ref{fig:boundary}(d). This suggests that the instability mechanism that we describe is also relevant for 
materials with anisotropic exchange. 

{\bf Conclusions}---We analyzed a new class of dynamic instability which is enabled by the topology of the order parameter manifold rather than kinematic effects. Such instability is an important mechanism that hinders the formation of spin superfluids. While the mechanism that we discuss is intrinsic to the Heisenberg model, an open problem is understanding the enhancement of the instability in the presence of localized defects, extended disorder and long-range interactions. In addition, extending our simulations to longer timescales in order to obtain a wholistic perspective of thermalization, which captures the growth of unstable modes and subsequent quasiparticle relaxation, is an important open challenge. 

{\bf Acknowledgements}---We thank Marin Bukov, Wen Wei Ho, Asier Pi\~{n}eiro Orioli, Daniel Podolsky, Ana Maria Rey, Amir Yacoby, Tony Zhou and Bihui Zhu for valuable insights and discussions. JFRN acknowledges the Gordon and Betty Moore Foundation’s EPiQS Initiative through Grant GBMF4302 and GBMF8686, the 2019 KITP program {\it Spin and Heat Transport in Quantum and Topological Materials}, and the National Science Foundation under Grant No. NSF PHY-1748958. AS and MK acknowledge support from the Technical University of Munich - Institute for Advanced Study, funded by the German Excellence Initiative and the European Union FP7 under grant agreement 291763, the Max Planck Gesellschaft (MPG) through the International Max Planck Research School for Quantum Science and Technology (IMPRS-QST), the Deutsche Forschungsgemeinschaft (DFG, German Research Foundation) under Germany's Excellence Strategy--EXC--2111--390814868, DFG TRR80 and DFG grant No. KN1254/1-2, and from the European Research Council (ERC) under the European Union's Horizon 2020 research and innovation programme (grant agreement No. 851161). ED acknowledges support from the Harvard-MIT CUA, AFOSR-MURI: Photonic Quantum Matter (award FA95501610323), and the DARPA DRINQS program (award D18AC00014).



\begin{thebibliography}{57}%
\makeatletter
\providecommand \@ifxundefined [1]{%
 \@ifx{#1\undefined}
}%
\providecommand \@ifnum [1]{%
 \ifnum #1\expandafter \@firstoftwo
 \else \expandafter \@secondoftwo
 \fi
}%
\providecommand \@ifx [1]{%
 \ifx #1\expandafter \@firstoftwo
 \else \expandafter \@secondoftwo
 \fi
}%
\providecommand \natexlab [1]{#1}%
\providecommand \enquote  [1]{``#1''}%
\providecommand \bibnamefont  [1]{#1}%
\providecommand \bibfnamefont [1]{#1}%
\providecommand \citenamefont [1]{#1}%
\providecommand \href@noop [0]{\@secondoftwo}%
\providecommand \href [0]{\begingroup \@sanitize@url \@href}%
\providecommand \@href[1]{\@@startlink{#1}\@@href}%
\providecommand \@@href[1]{\endgroup#1\@@endlink}%
\providecommand \@sanitize@url [0]{\catcode `\\12\catcode `\$12\catcode
  `\&12\catcode `\#12\catcode `\^12\catcode `\_12\catcode `\%12\relax}%
\providecommand \@@startlink[1]{}%
\providecommand \@@endlink[0]{}%
\providecommand \url  [0]{\begingroup\@sanitize@url \@url }%
\providecommand \@url [1]{\endgroup\@href {#1}{\urlprefix }}%
\providecommand \urlprefix  [0]{URL }%
\providecommand \Eprint [0]{\href }%
\providecommand \doibase [0]{http://dx.doi.org/}%
\providecommand \selectlanguage [0]{\@gobble}%
\providecommand \bibinfo  [0]{\@secondoftwo}%
\providecommand \bibfield  [0]{\@secondoftwo}%
\providecommand \translation [1]{[#1]}%
\providecommand \BibitemOpen [0]{}%
\providecommand \bibitemStop [0]{}%
\providecommand \bibitemNoStop [0]{.\EOS\space}%
\providecommand \EOS [0]{\spacefactor3000\relax}%
\providecommand \BibitemShut  [1]{\csname bibitem#1\endcsname}%
\let\auto@bib@innerbib\@empty
\bibitem [{\citenamefont {Landau}(1941)}]{1941landaucriterion}%
  \BibitemOpen
  \bibfield  {author} {\bibinfo {author} {\bibfnamefont {L.}~\bibnamefont
  {Landau}},\ }\href {\doibase 10.1103/PhysRev.60.356} {\bibfield  {journal}
  {\bibinfo  {journal} {Phys. Rev.}\ }\textbf {\bibinfo {volume} {60}},\
  \bibinfo {pages} {356} (\bibinfo {year} {1941})}\BibitemShut {NoStop}%
\bibitem [{\citenamefont {Dutton}\ \emph {et~al.}(2001)\citenamefont {Dutton},
  \citenamefont {Budde}, \citenamefont {Slowe},\ and\ \citenamefont
  {Hau}}]{2001sciencelandauinstability}%
  \BibitemOpen
  \bibfield  {author} {\bibinfo {author} {\bibfnamefont {Z.}~\bibnamefont
  {Dutton}}, \bibinfo {author} {\bibfnamefont {M.}~\bibnamefont {Budde}},
  \bibinfo {author} {\bibfnamefont {C.}~\bibnamefont {Slowe}}, \ and\ \bibinfo
  {author} {\bibfnamefont {L.~V.}\ \bibnamefont {Hau}},\ }\href {\doibase
  10.1126/science.1062527} {\bibfield  {journal} {\bibinfo  {journal}
  {Science}\ }\textbf {\bibinfo {volume} {293}},\ \bibinfo {pages} {663}
  (\bibinfo {year} {2001})}\BibitemShut {NoStop}%
\bibitem [{\citenamefont {Wu}\ and\ \citenamefont
  {Niu}(2001)}]{2001niukinematic}%
  \BibitemOpen
  \bibfield  {author} {\bibinfo {author} {\bibfnamefont {B.}~\bibnamefont
  {Wu}}\ and\ \bibinfo {author} {\bibfnamefont {Q.}~\bibnamefont {Niu}},\
  }\href {\doibase 10.1103/PhysRevA.64.061603} {\bibfield  {journal} {\bibinfo
  {journal} {Phys. Rev. A}\ }\textbf {\bibinfo {volume} {64}},\ \bibinfo
  {pages} {061603} (\bibinfo {year} {2001})}\BibitemShut {NoStop}%
\bibitem [{\citenamefont {Fallani}\ \emph {et~al.}(2004)\citenamefont
  {Fallani}, \citenamefont {De~Sarlo}, \citenamefont {Lye}, \citenamefont
  {Modugno}, \citenamefont {Saers}, \citenamefont {Fort},\ and\ \citenamefont
  {Inguscio}}]{2004prlmodulationalinstability}%
  \BibitemOpen
  \bibfield  {author} {\bibinfo {author} {\bibfnamefont {L.}~\bibnamefont
  {Fallani}}, \bibinfo {author} {\bibfnamefont {L.}~\bibnamefont {De~Sarlo}},
  \bibinfo {author} {\bibfnamefont {J.~E.}\ \bibnamefont {Lye}}, \bibinfo
  {author} {\bibfnamefont {M.}~\bibnamefont {Modugno}}, \bibinfo {author}
  {\bibfnamefont {R.}~\bibnamefont {Saers}}, \bibinfo {author} {\bibfnamefont
  {C.}~\bibnamefont {Fort}}, \ and\ \bibinfo {author} {\bibfnamefont
  {M.}~\bibnamefont {Inguscio}},\ }\href {\doibase
  10.1103/PhysRevLett.93.140406} {\bibfield  {journal} {\bibinfo  {journal}
  {Phys. Rev. Lett.}\ }\textbf {\bibinfo {volume} {93}},\ \bibinfo {pages}
  {140406} (\bibinfo {year} {2004})}\BibitemShut {NoStop}%
\bibitem [{\citenamefont {Polkovnikov}\ \emph {et~al.}(2005)\citenamefont
  {Polkovnikov}, \citenamefont {Altman}, \citenamefont {Demler}, \citenamefont
  {Halperin},\ and\ \citenamefont {Lukin}}]{2004polkovnikovmott}%
  \BibitemOpen
  \bibfield  {author} {\bibinfo {author} {\bibfnamefont {A.}~\bibnamefont
  {Polkovnikov}}, \bibinfo {author} {\bibfnamefont {E.}~\bibnamefont {Altman}},
  \bibinfo {author} {\bibfnamefont {E.}~\bibnamefont {Demler}}, \bibinfo
  {author} {\bibfnamefont {B.}~\bibnamefont {Halperin}}, \ and\ \bibinfo
  {author} {\bibfnamefont {M.~D.}\ \bibnamefont {Lukin}},\ }\href {\doibase
  10.1103/PhysRevA.71.063613} {\bibfield  {journal} {\bibinfo  {journal} {Phys.
  Rev. A}\ }\textbf {\bibinfo {volume} {71}},\ \bibinfo {pages} {063613}
  (\bibinfo {year} {2005})}\BibitemShut {NoStop}%
\bibitem [{\citenamefont {Altman}\ \emph {et~al.}(2005)\citenamefont {Altman},
  \citenamefont {Polkovnikov}, \citenamefont {Demler}, \citenamefont
  {Halperin},\ and\ \citenamefont {Lukin}}]{2005altmanmott}%
  \BibitemOpen
  \bibfield  {author} {\bibinfo {author} {\bibfnamefont {E.}~\bibnamefont
  {Altman}}, \bibinfo {author} {\bibfnamefont {A.}~\bibnamefont {Polkovnikov}},
  \bibinfo {author} {\bibfnamefont {E.}~\bibnamefont {Demler}}, \bibinfo
  {author} {\bibfnamefont {B.~I.}\ \bibnamefont {Halperin}}, \ and\ \bibinfo
  {author} {\bibfnamefont {M.~D.}\ \bibnamefont {Lukin}},\ }\href {\doibase
  10.1103/PhysRevLett.95.020402} {\bibfield  {journal} {\bibinfo  {journal}
  {Phys. Rev. Lett.}\ }\textbf {\bibinfo {volume} {95}},\ \bibinfo {pages}
  {020402} (\bibinfo {year} {2005})}\BibitemShut {NoStop}%
\bibitem [{\citenamefont {Norrie}\ \emph {et~al.}(2005)\citenamefont {Norrie},
  \citenamefont {Ballagh},\ and\ \citenamefont {Gardiner}}]{2005prlgardiner}%
  \BibitemOpen
  \bibfield  {author} {\bibinfo {author} {\bibfnamefont {A.~A.}\ \bibnamefont
  {Norrie}}, \bibinfo {author} {\bibfnamefont {R.~J.}\ \bibnamefont {Ballagh}},
  \ and\ \bibinfo {author} {\bibfnamefont {C.~W.}\ \bibnamefont {Gardiner}},\
  }\href {\doibase 10.1103/PhysRevLett.94.040401} {\bibfield  {journal}
  {\bibinfo  {journal} {Phys. Rev. Lett.}\ }\textbf {\bibinfo {volume} {94}},\
  \bibinfo {pages} {040401} (\bibinfo {year} {2005})}\BibitemShut {NoStop}%
\bibitem [{\citenamefont {Takeuchi}\ \emph {et~al.}(2010)\citenamefont
  {Takeuchi}, \citenamefont {Ishino},\ and\ \citenamefont
  {Tsubota}}]{2010prltsubota}%
  \BibitemOpen
  \bibfield  {author} {\bibinfo {author} {\bibfnamefont {H.}~\bibnamefont
  {Takeuchi}}, \bibinfo {author} {\bibfnamefont {S.}~\bibnamefont {Ishino}}, \
  and\ \bibinfo {author} {\bibfnamefont {M.}~\bibnamefont {Tsubota}},\ }\href
  {\doibase 10.1103/PhysRevLett.105.205301} {\bibfield  {journal} {\bibinfo
  {journal} {Phys. Rev. Lett.}\ }\textbf {\bibinfo {volume} {105}},\ \bibinfo
  {pages} {205301} (\bibinfo {year} {2010})}\BibitemShut {NoStop}%
\bibitem [{\citenamefont {Cherng}\ \emph {et~al.}(2008)\citenamefont {Cherng},
  \citenamefont {Gritsev}, \citenamefont {Stamper-Kurn},\ and\ \citenamefont
  {Demler}}]{2008instabilityxy}%
  \BibitemOpen
  \bibfield  {author} {\bibinfo {author} {\bibfnamefont {R.~W.}\ \bibnamefont
  {Cherng}}, \bibinfo {author} {\bibfnamefont {V.}~\bibnamefont {Gritsev}},
  \bibinfo {author} {\bibfnamefont {D.~M.}\ \bibnamefont {Stamper-Kurn}}, \
  and\ \bibinfo {author} {\bibfnamefont {E.}~\bibnamefont {Demler}},\ }\href
  {\doibase 10.1103/PhysRevLett.100.180404} {\bibfield  {journal} {\bibinfo
  {journal} {Phys. Rev. Lett.}\ }\textbf {\bibinfo {volume} {100}},\ \bibinfo
  {pages} {180404} (\bibinfo {year} {2008})}\BibitemShut {NoStop}%
\bibitem [{\citenamefont {Cherng}\ and\ \citenamefont
  {Demler}(2009)}]{2009spinorbec}%
  \BibitemOpen
  \bibfield  {author} {\bibinfo {author} {\bibfnamefont {R.~W.}\ \bibnamefont
  {Cherng}}\ and\ \bibinfo {author} {\bibfnamefont {E.}~\bibnamefont
  {Demler}},\ }\href {\doibase 10.1103/PhysRevLett.103.185301} {\bibfield
  {journal} {\bibinfo  {journal} {Phys. Rev. Lett.}\ }\textbf {\bibinfo
  {volume} {103}},\ \bibinfo {pages} {185301} (\bibinfo {year}
  {2009})}\BibitemShut {NoStop}%
\bibitem [{\citenamefont {Fujimoto}\ and\ \citenamefont
  {Tsubota}(2012)}]{2012pratsubota1}%
  \BibitemOpen
  \bibfield  {author} {\bibinfo {author} {\bibfnamefont {K.}~\bibnamefont
  {Fujimoto}}\ and\ \bibinfo {author} {\bibfnamefont {M.}~\bibnamefont
  {Tsubota}},\ }\href {\doibase 10.1103/PhysRevA.85.033642} {\bibfield
  {journal} {\bibinfo  {journal} {Phys. Rev. A}\ }\textbf {\bibinfo {volume}
  {85}},\ \bibinfo {pages} {033642} (\bibinfo {year} {2012})}\BibitemShut
  {NoStop}%
\bibitem [{\citenamefont {Langer}\ and\ \citenamefont
  {Ambegaokar}(1967)}]{1967scdecay}%
  \BibitemOpen
  \bibfield  {author} {\bibinfo {author} {\bibfnamefont {J.~S.}\ \bibnamefont
  {Langer}}\ and\ \bibinfo {author} {\bibfnamefont {V.}~\bibnamefont
  {Ambegaokar}},\ }\href {\doibase 10.1103/PhysRev.164.498} {\bibfield
  {journal} {\bibinfo  {journal} {Phys. Rev.}\ }\textbf {\bibinfo {volume}
  {164}},\ \bibinfo {pages} {498} (\bibinfo {year} {1967})}\BibitemShut
  {NoStop}%
\bibitem [{\citenamefont {McCumber}\ and\ \citenamefont
  {Halperin}(1970)}]{1970scdecay}%
  \BibitemOpen
  \bibfield  {author} {\bibinfo {author} {\bibfnamefont {D.~E.}\ \bibnamefont
  {McCumber}}\ and\ \bibinfo {author} {\bibfnamefont {B.~I.}\ \bibnamefont
  {Halperin}},\ }\href {\doibase 10.1103/PhysRevB.1.1054} {\bibfield  {journal}
  {\bibinfo  {journal} {Phys. Rev. B}\ }\textbf {\bibinfo {volume} {1}},\
  \bibinfo {pages} {1054} (\bibinfo {year} {1970})}\BibitemShut {NoStop}%
\bibitem [{\citenamefont {Sheehy}\ and\ \citenamefont
  {Goldbart}(1998)}]{1998so5scdecay}%
  \BibitemOpen
  \bibfield  {author} {\bibinfo {author} {\bibfnamefont {D.~E.}\ \bibnamefont
  {Sheehy}}\ and\ \bibinfo {author} {\bibfnamefont {P.~M.}\ \bibnamefont
  {Goldbart}},\ }\href {\doibase 10.1103/PhysRevB.57.R8131} {\bibfield
  {journal} {\bibinfo  {journal} {Phys. Rev. B}\ }\textbf {\bibinfo {volume}
  {57}},\ \bibinfo {pages} {R8131} (\bibinfo {year} {1998})}\BibitemShut
  {NoStop}%
\bibitem [{\citenamefont {Halperin}\ \emph {et~al.}(2010)\citenamefont
  {Halperin}, \citenamefont {Refael},\ and\ \citenamefont
  {Demler}}]{2010scdecayreview}%
  \BibitemOpen
  \bibfield  {author} {\bibinfo {author} {\bibfnamefont {B.~I.}\ \bibnamefont
  {Halperin}}, \bibinfo {author} {\bibfnamefont {G.}~\bibnamefont {Refael}}, \
  and\ \bibinfo {author} {\bibfnamefont {E.}~\bibnamefont {Demler}},\ }\href
  {\doibase 10.1142/S021797921005644X} {\bibfield  {journal} {\bibinfo
  {journal} {International Journal of Modern Physics B}\ }\textbf {\bibinfo
  {volume} {24}},\ \bibinfo {pages} {4039} (\bibinfo {year}
  {2010})}\BibitemShut {NoStop}%
\bibitem [{\citenamefont {Paul}\ \emph {et~al.}(2007)\citenamefont {Paul},
  \citenamefont {Schlagheck}, \citenamefont {Leboeuf},\ and\ \citenamefont
  {Pavloff}}]{2007becdisorder}%
  \BibitemOpen
  \bibfield  {author} {\bibinfo {author} {\bibfnamefont {T.}~\bibnamefont
  {Paul}}, \bibinfo {author} {\bibfnamefont {P.}~\bibnamefont {Schlagheck}},
  \bibinfo {author} {\bibfnamefont {P.}~\bibnamefont {Leboeuf}}, \ and\
  \bibinfo {author} {\bibfnamefont {N.}~\bibnamefont {Pavloff}},\ }\href
  {\doibase 10.1103/PhysRevLett.98.210602} {\bibfield  {journal} {\bibinfo
  {journal} {Phys. Rev. Lett.}\ }\textbf {\bibinfo {volume} {98}},\ \bibinfo
  {pages} {210602} (\bibinfo {year} {2007})}\BibitemShut {NoStop}%
\bibitem [{\citenamefont {Albert}\ \emph {et~al.}(2008)\citenamefont {Albert},
  \citenamefont {Paul}, \citenamefont {Pavloff},\ and\ \citenamefont
  {Leboeuf}}]{2008becdisorder}%
  \BibitemOpen
  \bibfield  {author} {\bibinfo {author} {\bibfnamefont {M.}~\bibnamefont
  {Albert}}, \bibinfo {author} {\bibfnamefont {T.}~\bibnamefont {Paul}},
  \bibinfo {author} {\bibfnamefont {N.}~\bibnamefont {Pavloff}}, \ and\
  \bibinfo {author} {\bibfnamefont {P.}~\bibnamefont {Leboeuf}},\ }\href
  {\doibase 10.1103/PhysRevLett.100.250405} {\bibfield  {journal} {\bibinfo
  {journal} {Phys. Rev. Lett.}\ }\textbf {\bibinfo {volume} {100}},\ \bibinfo
  {pages} {250405} (\bibinfo {year} {2008})}\BibitemShut {NoStop}%
\bibitem [{\citenamefont {Albert}\ \emph {et~al.}(2010)\citenamefont {Albert},
  \citenamefont {Paul}, \citenamefont {Pavloff},\ and\ \citenamefont
  {Leboeuf}}]{2010becdisorder}%
  \BibitemOpen
  \bibfield  {author} {\bibinfo {author} {\bibfnamefont {M.}~\bibnamefont
  {Albert}}, \bibinfo {author} {\bibfnamefont {T.}~\bibnamefont {Paul}},
  \bibinfo {author} {\bibfnamefont {N.}~\bibnamefont {Pavloff}}, \ and\
  \bibinfo {author} {\bibfnamefont {P.}~\bibnamefont {Leboeuf}},\ }\href
  {\doibase 10.1103/PhysRevA.82.011602} {\bibfield  {journal} {\bibinfo
  {journal} {Phys. Rev. A}\ }\textbf {\bibinfo {volume} {82}},\ \bibinfo
  {pages} {011602} (\bibinfo {year} {2010})}\BibitemShut {NoStop}%
\bibitem [{\citenamefont {Haga}\ and\ \citenamefont {Ueda}(2019)}]{2019ueda}%
  \BibitemOpen
  \bibfield  {author} {\bibinfo {author} {\bibfnamefont {T.}~\bibnamefont
  {Haga}}\ and\ \bibinfo {author} {\bibfnamefont {M.}~\bibnamefont {Ueda}},\
  }\href@noop {} {\enquote {\bibinfo {title} {Anomalous phase fluctuations of a
  superfluid flowing in a random potential},}\ } (\bibinfo {year} {2019}),\
  \Eprint {http://arxiv.org/abs/arXiv:1909.11997} {arXiv:arXiv:1909.11997
  [cond-mat.quant-gas]} \BibitemShut {NoStop}%
\bibitem [{\citenamefont {Santos}\ \emph {et~al.}(2000)\citenamefont {Santos},
  \citenamefont {Shlyapnikov}, \citenamefont {Zoller},\ and\ \citenamefont
  {Lewenstein}}]{2000dipolarbec}%
  \BibitemOpen
  \bibfield  {author} {\bibinfo {author} {\bibfnamefont {L.}~\bibnamefont
  {Santos}}, \bibinfo {author} {\bibfnamefont {G.~V.}\ \bibnamefont
  {Shlyapnikov}}, \bibinfo {author} {\bibfnamefont {P.}~\bibnamefont {Zoller}},
  \ and\ \bibinfo {author} {\bibfnamefont {M.}~\bibnamefont {Lewenstein}},\
  }\href {\doibase 10.1103/PhysRevLett.85.1791} {\bibfield  {journal} {\bibinfo
   {journal} {Phys. Rev. Lett.}\ }\textbf {\bibinfo {volume} {85}},\ \bibinfo
  {pages} {1791} (\bibinfo {year} {2000})}\BibitemShut {NoStop}%
\bibitem [{\citenamefont {G\'oral}\ and\ \citenamefont
  {Santos}(2002)}]{2002dipolarbec}%
  \BibitemOpen
  \bibfield  {author} {\bibinfo {author} {\bibfnamefont {K.}~\bibnamefont
  {G\'oral}}\ and\ \bibinfo {author} {\bibfnamefont {L.}~\bibnamefont
  {Santos}},\ }\href {\doibase 10.1103/PhysRevA.66.023613} {\bibfield
  {journal} {\bibinfo  {journal} {Phys. Rev. A}\ }\textbf {\bibinfo {volume}
  {66}},\ \bibinfo {pages} {023613} (\bibinfo {year} {2002})}\BibitemShut
  {NoStop}%
\bibitem [{\citenamefont {Santos}\ \emph {et~al.}(2003)\citenamefont {Santos},
  \citenamefont {Shlyapnikov},\ and\ \citenamefont
  {Lewenstein}}]{2003dipolarbec}%
  \BibitemOpen
  \bibfield  {author} {\bibinfo {author} {\bibfnamefont {L.}~\bibnamefont
  {Santos}}, \bibinfo {author} {\bibfnamefont {G.~V.}\ \bibnamefont
  {Shlyapnikov}}, \ and\ \bibinfo {author} {\bibfnamefont {M.}~\bibnamefont
  {Lewenstein}},\ }\href {\doibase 10.1103/PhysRevLett.90.250403} {\bibfield
  {journal} {\bibinfo  {journal} {Phys. Rev. Lett.}\ }\textbf {\bibinfo
  {volume} {90}},\ \bibinfo {pages} {250403} (\bibinfo {year}
  {2003})}\BibitemShut {NoStop}%
\bibitem [{\citenamefont {Bertotti}\ \emph {et~al.}(2001)\citenamefont
  {Bertotti}, \citenamefont {Mayergoyz},\ and\ \citenamefont
  {Serpico}}]{2001parametricpumping}%
  \BibitemOpen
  \bibfield  {author} {\bibinfo {author} {\bibfnamefont {G.}~\bibnamefont
  {Bertotti}}, \bibinfo {author} {\bibfnamefont {I.}~\bibnamefont {Mayergoyz}},
  \ and\ \bibinfo {author} {\bibfnamefont {C.}~\bibnamefont {Serpico}},\ }\href
  {\doibase https://doi.org/10.1016/S0921-4526(01)00987-5} {\bibfield
  {journal} {\bibinfo  {journal} {Physica B: Condensed Matter}\ }\textbf
  {\bibinfo {volume} {306}},\ \bibinfo {pages} {106 } (\bibinfo {year}
  {2001})},\ \bibinfo {note} {proceedings of the Third International Symposium
  on Hysteresis an d Micromagnetics Modeling}\BibitemShut {NoStop}%
\bibitem [{\citenamefont {Kreil}\ \emph {et~al.}(2018)\citenamefont {Kreil},
  \citenamefont {Bozhko}, \citenamefont {Musiienko-Shmarova}, \citenamefont
  {Vasyuchka}, \citenamefont {L'vov}, \citenamefont {Pomyalov}, \citenamefont
  {Hillebrands},\ and\ \citenamefont {Serga}}]{2018magnoninstabilities}%
  \BibitemOpen
  \bibfield  {author} {\bibinfo {author} {\bibfnamefont {A.~J.~E.}\
  \bibnamefont {Kreil}}, \bibinfo {author} {\bibfnamefont {D.~A.}\ \bibnamefont
  {Bozhko}}, \bibinfo {author} {\bibfnamefont {H.~Y.}\ \bibnamefont
  {Musiienko-Shmarova}}, \bibinfo {author} {\bibfnamefont {V.~I.}\ \bibnamefont
  {Vasyuchka}}, \bibinfo {author} {\bibfnamefont {V.~S.}\ \bibnamefont
  {L'vov}}, \bibinfo {author} {\bibfnamefont {A.}~\bibnamefont {Pomyalov}},
  \bibinfo {author} {\bibfnamefont {B.}~\bibnamefont {Hillebrands}}, \ and\
  \bibinfo {author} {\bibfnamefont {A.~A.}\ \bibnamefont {Serga}},\ }\href
  {\doibase 10.1103/PhysRevLett.121.077203} {\bibfield  {journal} {\bibinfo
  {journal} {Phys. Rev. Lett.}\ }\textbf {\bibinfo {volume} {121}},\ \bibinfo
  {pages} {077203} (\bibinfo {year} {2018})}\BibitemShut {NoStop}%
\bibitem [{\citenamefont {Wintersperger}\ \emph {et~al.}(2018)\citenamefont
  {Wintersperger}, \citenamefont {Bukov}, \citenamefont {Näger}, \citenamefont
  {Lellouch}, \citenamefont {Demler}, \citenamefont {Schneider}, \citenamefont
  {Bloch}, \citenamefont {Goldman},\ and\ \citenamefont
  {Aidelsburger}}]{2018heatingdrivenbec}%
  \BibitemOpen
  \bibfield  {author} {\bibinfo {author} {\bibfnamefont {K.}~\bibnamefont
  {Wintersperger}}, \bibinfo {author} {\bibfnamefont {M.}~\bibnamefont
  {Bukov}}, \bibinfo {author} {\bibfnamefont {J.}~\bibnamefont {Näger}},
  \bibinfo {author} {\bibfnamefont {S.}~\bibnamefont {Lellouch}}, \bibinfo
  {author} {\bibfnamefont {E.}~\bibnamefont {Demler}}, \bibinfo {author}
  {\bibfnamefont {U.}~\bibnamefont {Schneider}}, \bibinfo {author}
  {\bibfnamefont {I.}~\bibnamefont {Bloch}}, \bibinfo {author} {\bibfnamefont
  {N.}~\bibnamefont {Goldman}}, \ and\ \bibinfo {author} {\bibfnamefont
  {M.}~\bibnamefont {Aidelsburger}},\ }\href@noop {} {\enquote {\bibinfo
  {title} {Parametric instabilities of interacting bosons in
  periodically-driven 1d optical lattices},}\ } (\bibinfo {year} {2018}),\
  \Eprint {http://arxiv.org/abs/1808.07462} {arXiv:1808.07462
  [cond-mat.quant-gas]} \BibitemShut {NoStop}%
\bibitem [{\citenamefont {Bukov}\ \emph {et~al.}(2015)\citenamefont {Bukov},
  \citenamefont {Gopalakrishnan}, \citenamefont {Knap},\ and\ \citenamefont
  {Demler}}]{2015bukovfloquet}%
  \BibitemOpen
  \bibfield  {author} {\bibinfo {author} {\bibfnamefont {M.}~\bibnamefont
  {Bukov}}, \bibinfo {author} {\bibfnamefont {S.}~\bibnamefont
  {Gopalakrishnan}}, \bibinfo {author} {\bibfnamefont {M.}~\bibnamefont
  {Knap}}, \ and\ \bibinfo {author} {\bibfnamefont {E.}~\bibnamefont
  {Demler}},\ }\href {\doibase 10.1103/PhysRevLett.115.205301} {\bibfield
  {journal} {\bibinfo  {journal} {Phys. Rev. Lett.}\ }\textbf {\bibinfo
  {volume} {115}},\ \bibinfo {pages} {205301} (\bibinfo {year}
  {2015})}\BibitemShut {NoStop}%
\bibitem [{\citenamefont {Boulier}\ \emph {et~al.}(2019)\citenamefont
  {Boulier}, \citenamefont {Maslek}, \citenamefont {Bukov}, \citenamefont
  {Bracamontes}, \citenamefont {Magnan}, \citenamefont {Lellouch},
  \citenamefont {Demler}, \citenamefont {Goldman},\ and\ \citenamefont
  {Porto}}]{2019instabilitydrivenbec}%
  \BibitemOpen
  \bibfield  {author} {\bibinfo {author} {\bibfnamefont {T.}~\bibnamefont
  {Boulier}}, \bibinfo {author} {\bibfnamefont {J.}~\bibnamefont {Maslek}},
  \bibinfo {author} {\bibfnamefont {M.}~\bibnamefont {Bukov}}, \bibinfo
  {author} {\bibfnamefont {C.}~\bibnamefont {Bracamontes}}, \bibinfo {author}
  {\bibfnamefont {E.}~\bibnamefont {Magnan}}, \bibinfo {author} {\bibfnamefont
  {S.}~\bibnamefont {Lellouch}}, \bibinfo {author} {\bibfnamefont
  {E.}~\bibnamefont {Demler}}, \bibinfo {author} {\bibfnamefont
  {N.}~\bibnamefont {Goldman}}, \ and\ \bibinfo {author} {\bibfnamefont
  {J.~V.}\ \bibnamefont {Porto}},\ }\href {\doibase 10.1103/PhysRevX.9.011047}
  {\bibfield  {journal} {\bibinfo  {journal} {Phys. Rev. X}\ }\textbf {\bibinfo
  {volume} {9}},\ \bibinfo {pages} {011047} (\bibinfo {year}
  {2019})}\BibitemShut {NoStop}%
\bibitem [{\citenamefont {Bardon}\ \emph {et~al.}(2014)\citenamefont {Bardon},
  \citenamefont {Beattie}, \citenamefont {Luciuk}, \citenamefont {Cairncross},
  \citenamefont {Fine}, \citenamefont {Cheng}, \citenamefont {Edge},
  \citenamefont {Taylor}, \citenamefont {Zhang}, \citenamefont {Trotzky},\ and\
  \citenamefont {Thywissen}}]{2014spiralscience}%
  \BibitemOpen
  \bibfield  {author} {\bibinfo {author} {\bibfnamefont {A.~B.}\ \bibnamefont
  {Bardon}}, \bibinfo {author} {\bibfnamefont {S.}~\bibnamefont {Beattie}},
  \bibinfo {author} {\bibfnamefont {C.}~\bibnamefont {Luciuk}}, \bibinfo
  {author} {\bibfnamefont {W.}~\bibnamefont {Cairncross}}, \bibinfo {author}
  {\bibfnamefont {D.}~\bibnamefont {Fine}}, \bibinfo {author} {\bibfnamefont
  {N.~S.}\ \bibnamefont {Cheng}}, \bibinfo {author} {\bibfnamefont {G.~J.~A.}\
  \bibnamefont {Edge}}, \bibinfo {author} {\bibfnamefont {E.}~\bibnamefont
  {Taylor}}, \bibinfo {author} {\bibfnamefont {S.}~\bibnamefont {Zhang}},
  \bibinfo {author} {\bibfnamefont {S.}~\bibnamefont {Trotzky}}, \ and\
  \bibinfo {author} {\bibfnamefont {J.~H.}\ \bibnamefont {Thywissen}},\ }\href
  {\doibase 10.1126/science.1247425} {\bibfield  {journal} {\bibinfo  {journal}
  {Science}\ }\textbf {\bibinfo {volume} {344}},\ \bibinfo {pages} {722}
  (\bibinfo {year} {2014})}\BibitemShut {NoStop}%
\bibitem [{\citenamefont {Hild}\ \emph {et~al.}(2014)\citenamefont {Hild},
  \citenamefont {Fukuhara}, \citenamefont {Schau\ss{}}, \citenamefont {Zeiher},
  \citenamefont {Knap}, \citenamefont {Demler}, \citenamefont {Bloch},\ and\
  \citenamefont {Gross}}]{2014spiralexp}%
  \BibitemOpen
  \bibfield  {author} {\bibinfo {author} {\bibfnamefont {S.}~\bibnamefont
  {Hild}}, \bibinfo {author} {\bibfnamefont {T.}~\bibnamefont {Fukuhara}},
  \bibinfo {author} {\bibfnamefont {P.}~\bibnamefont {Schau\ss{}}}, \bibinfo
  {author} {\bibfnamefont {J.}~\bibnamefont {Zeiher}}, \bibinfo {author}
  {\bibfnamefont {M.}~\bibnamefont {Knap}}, \bibinfo {author} {\bibfnamefont
  {E.}~\bibnamefont {Demler}}, \bibinfo {author} {\bibfnamefont
  {I.}~\bibnamefont {Bloch}}, \ and\ \bibinfo {author} {\bibfnamefont
  {C.}~\bibnamefont {Gross}},\ }\href {\doibase 10.1103/PhysRevLett.113.147205}
  {\bibfield  {journal} {\bibinfo  {journal} {Phys. Rev. Lett.}\ }\textbf
  {\bibinfo {volume} {113}},\ \bibinfo {pages} {147205} (\bibinfo {year}
  {2014})}\BibitemShut {NoStop}%
\bibitem [{\citenamefont {Brown}\ \emph {et~al.}(2015)\citenamefont {Brown},
  \citenamefont {Wyllie}, \citenamefont {Koller}, \citenamefont {Goldschmidt},
  \citenamefont {Foss-Feig},\ and\ \citenamefont {Porto}}]{2015sciencespiral}%
  \BibitemOpen
  \bibfield  {author} {\bibinfo {author} {\bibfnamefont {R.~C.}\ \bibnamefont
  {Brown}}, \bibinfo {author} {\bibfnamefont {R.}~\bibnamefont {Wyllie}},
  \bibinfo {author} {\bibfnamefont {S.~B.}\ \bibnamefont {Koller}}, \bibinfo
  {author} {\bibfnamefont {E.~A.}\ \bibnamefont {Goldschmidt}}, \bibinfo
  {author} {\bibfnamefont {M.}~\bibnamefont {Foss-Feig}}, \ and\ \bibinfo
  {author} {\bibfnamefont {J.~V.}\ \bibnamefont {Porto}},\ }\href {\doibase
  10.1126/science.aaa1385} {\bibfield  {journal} {\bibinfo  {journal}
  {Science}\ }\textbf {\bibinfo {volume} {348}},\ \bibinfo {pages} {540}
  (\bibinfo {year} {2015})}\BibitemShut {NoStop}%
\bibitem [{\citenamefont {Tsubota}(2009)}]{2009tsubotareview}%
  \BibitemOpen
  \bibfield  {author} {\bibinfo {author} {\bibfnamefont {M.}~\bibnamefont
  {Tsubota}},\ }\href {\doibase 10.1088/0953-8984/21/16/164207} {\bibfield
  {journal} {\bibinfo  {journal} {Journal of Physics: Condensed Matter}\
  }\textbf {\bibinfo {volume} {21}},\ \bibinfo {pages} {164207} (\bibinfo
  {year} {2009})}\BibitemShut {NoStop}%
\bibitem [{\citenamefont {Rodriguez-Nieva}(2020)}]{2020turbulence}%
  \BibitemOpen
  \bibfield  {author} {\bibinfo {author} {\bibfnamefont {J.~F.}\ \bibnamefont
  {Rodriguez-Nieva}},\ }\href@noop {} {\enquote {\bibinfo {title} {Turbulent
  relaxation after a quench in the heisenberg model},}\ } (\bibinfo {year}
  {2020}),\ \Eprint {http://arxiv.org/abs/arXiv:2009.11883}
  {arXiv:arXiv:2009.11883} \BibitemShut {NoStop}%
\bibitem [{\citenamefont {Babadi}\ \emph {et~al.}(2015)\citenamefont {Babadi},
  \citenamefont {Demler},\ and\ \citenamefont {Knap}}]{2015PRX-babadi}%
  \BibitemOpen
  \bibfield  {author} {\bibinfo {author} {\bibfnamefont {M.}~\bibnamefont
  {Babadi}}, \bibinfo {author} {\bibfnamefont {E.}~\bibnamefont {Demler}}, \
  and\ \bibinfo {author} {\bibfnamefont {M.}~\bibnamefont {Knap}},\ }\href
  {\doibase 10.1103/PhysRevX.5.041005} {\bibfield  {journal} {\bibinfo
  {journal} {Phys. Rev. X}\ }\textbf {\bibinfo {volume} {5}},\ \bibinfo {pages}
  {041005} (\bibinfo {year} {2015})}\BibitemShut {NoStop}%
\bibitem [{\citenamefont {Pi\~neiro Orioli}\ \emph {et~al.}(2015)\citenamefont
  {Pi\~neiro Orioli}, \citenamefont {Boguslavski},\ and\ \citenamefont
  {Berges}}]{2015asier}%
  \BibitemOpen
  \bibfield  {author} {\bibinfo {author} {\bibfnamefont {A.}~\bibnamefont
  {Pi\~neiro Orioli}}, \bibinfo {author} {\bibfnamefont {K.}~\bibnamefont
  {Boguslavski}}, \ and\ \bibinfo {author} {\bibfnamefont {J.}~\bibnamefont
  {Berges}},\ }\href {\doibase 10.1103/PhysRevD.92.025041} {\bibfield
  {journal} {\bibinfo  {journal} {Phys. Rev. D}\ }\textbf {\bibinfo {volume}
  {92}},\ \bibinfo {pages} {025041} (\bibinfo {year} {2015})}\BibitemShut
  {NoStop}%
\bibitem [{\citenamefont {{Berges}}(2015)}]{2015bergesreview}%
  \BibitemOpen
  \bibfield  {author} {\bibinfo {author} {\bibfnamefont {J.}~\bibnamefont
  {{Berges}}},\ }\href@noop {} {\bibfield  {journal} {\bibinfo  {journal}
  {arXiv e-prints}\ ,\ \bibinfo {eid} {arXiv:1503.02907}} (\bibinfo {year}
  {2015})}\BibitemShut {NoStop}%
\bibitem [{\citenamefont {Jepsen}\ \emph {et~al.}(2020)\citenamefont {Jepsen},
  \citenamefont {Amato-Grill}, \citenamefont {Dimitrova}, \citenamefont {Ho},
  \citenamefont {Demler},\ and\ \citenamefont {Ketterle}}]{2020spiralketterle}%
  \BibitemOpen
  \bibfield  {author} {\bibinfo {author} {\bibfnamefont {N.}~\bibnamefont
  {Jepsen}}, \bibinfo {author} {\bibfnamefont {J.}~\bibnamefont {Amato-Grill}},
  \bibinfo {author} {\bibfnamefont {I.}~\bibnamefont {Dimitrova}}, \bibinfo
  {author} {\bibfnamefont {W.~W.}\ \bibnamefont {Ho}}, \bibinfo {author}
  {\bibfnamefont {E.}~\bibnamefont {Demler}}, \ and\ \bibinfo {author}
  {\bibfnamefont {W.}~\bibnamefont {Ketterle}},\ }\href@noop {} {\enquote
  {\bibinfo {title} {Spin transport in a tunable heisenberg model realized with
  ultracold atoms},}\ } (\bibinfo {year} {2020}),\ \Eprint
  {http://arxiv.org/abs/2005.09549} {arXiv:2005.09549 [cond-mat.quant-gas]}
  \BibitemShut {NoStop}%
\bibitem [{\citenamefont {Sonin}(2010)}]{spinsuperfluidreview}%
  \BibitemOpen
  \bibfield  {author} {\bibinfo {author} {\bibfnamefont {E.}~\bibnamefont
  {Sonin}},\ }\href {\doibase 10.1080/00018731003739943} {\bibfield  {journal}
  {\bibinfo  {journal} {Advances in Physics}\ }\textbf {\bibinfo {volume}
  {59}},\ \bibinfo {pages} {181} (\bibinfo {year} {2010})}\BibitemShut
  {NoStop}%
\bibitem [{\citenamefont {Bozhko}\ \emph
  {et~al.}(2016{\natexlab{a}})\citenamefont {Bozhko}, \citenamefont {Serga},
  \citenamefont {Clausen}, \citenamefont {Vasyuchka}, \citenamefont {Heussner},
  \citenamefont {Melkov}, \citenamefont {Pomyalov}, \citenamefont {L’vov},\
  and\ \citenamefont {Hillebrands}}]{2016supercurrentyig}%
  \BibitemOpen
  \bibfield  {author} {\bibinfo {author} {\bibfnamefont {D.~A.}\ \bibnamefont
  {Bozhko}}, \bibinfo {author} {\bibfnamefont {A.~A.}\ \bibnamefont {Serga}},
  \bibinfo {author} {\bibfnamefont {P.}~\bibnamefont {Clausen}}, \bibinfo
  {author} {\bibfnamefont {V.~I.}\ \bibnamefont {Vasyuchka}}, \bibinfo {author}
  {\bibfnamefont {F.}~\bibnamefont {Heussner}}, \bibinfo {author}
  {\bibfnamefont {G.~A.}\ \bibnamefont {Melkov}}, \bibinfo {author}
  {\bibfnamefont {A.}~\bibnamefont {Pomyalov}}, \bibinfo {author}
  {\bibfnamefont {V.~S.}\ \bibnamefont {L’vov}}, \ and\ \bibinfo {author}
  {\bibfnamefont {B.}~\bibnamefont {Hillebrands}},\ }\href
  {https://doi.org/10.1038/nphys3838} {\bibfield  {journal} {\bibinfo
  {journal} {Nature Physics}\ }\textbf {\bibinfo {volume} {12}},\ \bibinfo
  {pages} {1057} (\bibinfo {year} {2016}{\natexlab{a}})}\BibitemShut {NoStop}%
\bibitem [{\citenamefont {Sonin}(2016)}]{2016sonincomment}%
  \BibitemOpen
  \bibfield  {author} {\bibinfo {author} {\bibfnamefont {E.~B.}\ \bibnamefont
  {Sonin}},\ }\href@noop {} {\enquote {\bibinfo {title} {Comment on
  "supercurrent in a room temperature bose-einstein magnon condensate"},}\ }
  (\bibinfo {year} {2016}),\ \Eprint {http://arxiv.org/abs/1607.04720}
  {arXiv:1607.04720 [cond-mat.other]} \BibitemShut {NoStop}%
\bibitem [{\citenamefont {Bozhko}\ \emph
  {et~al.}(2016{\natexlab{b}})\citenamefont {Bozhko}, \citenamefont {Serga},
  \citenamefont {Clausen}, \citenamefont {Vasyuchka}, \citenamefont {Melkov},
  \citenamefont {L'vov},\ and\ \citenamefont
  {Hillebrands}}]{2016hillebrandsreply}%
  \BibitemOpen
  \bibfield  {author} {\bibinfo {author} {\bibfnamefont {D.~A.}\ \bibnamefont
  {Bozhko}}, \bibinfo {author} {\bibfnamefont {A.~A.}\ \bibnamefont {Serga}},
  \bibinfo {author} {\bibfnamefont {P.}~\bibnamefont {Clausen}}, \bibinfo
  {author} {\bibfnamefont {V.~I.}\ \bibnamefont {Vasyuchka}}, \bibinfo {author}
  {\bibfnamefont {G.~A.}\ \bibnamefont {Melkov}}, \bibinfo {author}
  {\bibfnamefont {V.~S.}\ \bibnamefont {L'vov}}, \ and\ \bibinfo {author}
  {\bibfnamefont {B.}~\bibnamefont {Hillebrands}},\ }\href@noop {} {\enquote
  {\bibinfo {title} {On supercurrents in bose-einstein magnon condensates in
  yig ferrimagnet},}\ } (\bibinfo {year} {2016}{\natexlab{b}}),\ \Eprint
  {http://arxiv.org/abs/1608.01813} {arXiv:1608.01813 [cond-mat.other]}
  \BibitemShut {NoStop}%
\bibitem [{\citenamefont {Takei}\ and\ \citenamefont
  {Tserkovnyak}(2014)}]{2014yaroslavsuperfluid}%
  \BibitemOpen
  \bibfield  {author} {\bibinfo {author} {\bibfnamefont {S.}~\bibnamefont
  {Takei}}\ and\ \bibinfo {author} {\bibfnamefont {Y.}~\bibnamefont
  {Tserkovnyak}},\ }\href {\doibase 10.1103/PhysRevLett.112.227201} {\bibfield
  {journal} {\bibinfo  {journal} {Phys. Rev. Lett.}\ }\textbf {\bibinfo
  {volume} {112}},\ \bibinfo {pages} {227201} (\bibinfo {year}
  {2014})}\BibitemShut {NoStop}%
\bibitem [{\citenamefont {Nakata}\ \emph {et~al.}(2014)\citenamefont {Nakata},
  \citenamefont {van Hoogdalem}, \citenamefont {Simon},\ and\ \citenamefont
  {Loss}}]{2014lossspinsuperfluidity}%
  \BibitemOpen
  \bibfield  {author} {\bibinfo {author} {\bibfnamefont {K.}~\bibnamefont
  {Nakata}}, \bibinfo {author} {\bibfnamefont {K.~A.}\ \bibnamefont {van
  Hoogdalem}}, \bibinfo {author} {\bibfnamefont {P.}~\bibnamefont {Simon}}, \
  and\ \bibinfo {author} {\bibfnamefont {D.}~\bibnamefont {Loss}},\ }\href
  {\doibase 10.1103/PhysRevB.90.144419} {\bibfield  {journal} {\bibinfo
  {journal} {Phys. Rev. B}\ }\textbf {\bibinfo {volume} {90}},\ \bibinfo
  {pages} {144419} (\bibinfo {year} {2014})}\BibitemShut {NoStop}%
\bibitem [{\citenamefont {Sun}\ \emph {et~al.}(2016)\citenamefont {Sun},
  \citenamefont {Nattermann},\ and\ \citenamefont
  {Pokrovsky}}]{2016yigsuperfluidity}%
  \BibitemOpen
  \bibfield  {author} {\bibinfo {author} {\bibfnamefont {C.}~\bibnamefont
  {Sun}}, \bibinfo {author} {\bibfnamefont {T.}~\bibnamefont {Nattermann}}, \
  and\ \bibinfo {author} {\bibfnamefont {V.~L.}\ \bibnamefont {Pokrovsky}},\
  }\href {\doibase 10.1103/PhysRevLett.116.257205} {\bibfield  {journal}
  {\bibinfo  {journal} {Phys. Rev. Lett.}\ }\textbf {\bibinfo {volume} {116}},\
  \bibinfo {pages} {257205} (\bibinfo {year} {2016})}\BibitemShut {NoStop}%
\bibitem [{\citenamefont {Sonin}(2017)}]{2017spinsuperfluiditysonin}%
  \BibitemOpen
  \bibfield  {author} {\bibinfo {author} {\bibfnamefont {E.~B.}\ \bibnamefont
  {Sonin}},\ }\href {\doibase 10.1103/PhysRevB.95.144432} {\bibfield  {journal}
  {\bibinfo  {journal} {Phys. Rev. B}\ }\textbf {\bibinfo {volume} {95}},\
  \bibinfo {pages} {144432} (\bibinfo {year} {2017})}\BibitemShut {NoStop}%
\bibitem [{\citenamefont {Mattis}(2006)}]{mattisbook}%
  \BibitemOpen
  \bibfield  {author} {\bibinfo {author} {\bibfnamefont {D.~C.}\ \bibnamefont
  {Mattis}},\ }\href {\doibase 10.1142/5372} {\emph {\bibinfo {title} {The
  Theory of Magnetism Made Simple}}}\ (\bibinfo  {publisher} {WORLD
  SCIENTIFIC},\ \bibinfo {year} {2006})\BibitemShut {NoStop}%
\bibitem [{\citenamefont {Bhattacharyya}\ \emph {et~al.}(2019)\citenamefont
  {Bhattacharyya}, \citenamefont {Rodriguez-Nieva},\ and\ \citenamefont
  {Demler}}]{2019saraswat}%
  \BibitemOpen
  \bibfield  {author} {\bibinfo {author} {\bibfnamefont {S.}~\bibnamefont
  {Bhattacharyya}}, \bibinfo {author} {\bibfnamefont {J.~F.}\ \bibnamefont
  {Rodriguez-Nieva}}, \ and\ \bibinfo {author} {\bibfnamefont {E.}~\bibnamefont
  {Demler}},\ }\href@noop {} {\enquote {\bibinfo {title} {Universal dynamics
  far from equilibrium in heisenberg ferromagnets},}\ } (\bibinfo {year}
  {2019}),\ \Eprint {http://arxiv.org/abs/1908.00554} {arXiv:1908.00554
  [cond-mat.stat-mech]} \BibitemShut {NoStop}%
\bibitem [{\citenamefont {Iacocca}\ \emph {et~al.}(2017)\citenamefont
  {Iacocca}, \citenamefont {Silva},\ and\ \citenamefont
  {Hoefer}}]{2017breakinggalilean}%
  \BibitemOpen
  \bibfield  {author} {\bibinfo {author} {\bibfnamefont {E.}~\bibnamefont
  {Iacocca}}, \bibinfo {author} {\bibfnamefont {T.~J.}\ \bibnamefont {Silva}},
  \ and\ \bibinfo {author} {\bibfnamefont {M.~A.}\ \bibnamefont {Hoefer}},\
  }\href {\doibase 10.1103/PhysRevLett.118.017203} {\bibfield  {journal}
  {\bibinfo  {journal} {Phys. Rev. Lett.}\ }\textbf {\bibinfo {volume} {118}},\
  \bibinfo {pages} {017203} (\bibinfo {year} {2017})}\BibitemShut {NoStop}%
\bibitem [{\citenamefont {Rodriguez-Nieva}\ \emph {et~al.}(2018)\citenamefont
  {Rodriguez-Nieva}, \citenamefont {Podolsky},\ and\ \citenamefont
  {Demler}}]{2018hydrodynamicsoundmodes}%
  \BibitemOpen
  \bibfield  {author} {\bibinfo {author} {\bibfnamefont {J.~F.}\ \bibnamefont
  {Rodriguez-Nieva}}, \bibinfo {author} {\bibfnamefont {D.}~\bibnamefont
  {Podolsky}}, \ and\ \bibinfo {author} {\bibfnamefont {E.}~\bibnamefont
  {Demler}},\ }\href@noop {} {\enquote {\bibinfo {title} {Hydrodynamic sound
  modes and galilean symmetry breaking in a magnon fluid},}\ } (\bibinfo {year}
  {2018}),\ \Eprint {http://arxiv.org/abs/1810.12333} {arXiv:1810.12333
  [cond-mat.mes-hall]} \BibitemShut {NoStop}%
\bibitem [{\citenamefont {Polkovnikov}(2010)}]{2010polkovnikovreview}%
  \BibitemOpen
  \bibfield  {author} {\bibinfo {author} {\bibfnamefont {A.}~\bibnamefont
  {Polkovnikov}},\ }\href {\doibase https://doi.org/10.1016/j.aop.2010.02.006}
  {\bibfield  {journal} {\bibinfo  {journal} {Annals of Physics}\ }\textbf
  {\bibinfo {volume} {325}},\ \bibinfo {pages} {1790 } (\bibinfo {year}
  {2010})}\BibitemShut {NoStop}%
\bibitem [{\citenamefont {Berges}\ and\ \citenamefont
  {Serreau}(2003)}]{2003prlberges}%
  \BibitemOpen
  \bibfield  {author} {\bibinfo {author} {\bibfnamefont {J.}~\bibnamefont
  {Berges}}\ and\ \bibinfo {author} {\bibfnamefont {J.}~\bibnamefont
  {Serreau}},\ }\href {\doibase 10.1103/PhysRevLett.91.111601} {\bibfield
  {journal} {\bibinfo  {journal} {Phys. Rev. Lett.}\ }\textbf {\bibinfo
  {volume} {91}},\ \bibinfo {pages} {111601} (\bibinfo {year}
  {2003})}\BibitemShut {NoStop}%
\bibitem [{\citenamefont {Berges}\ \emph {et~al.}(2008)\citenamefont {Berges},
  \citenamefont {Rothkopf},\ and\ \citenamefont {Schmidt}}]{2008berges}%
  \BibitemOpen
  \bibfield  {author} {\bibinfo {author} {\bibfnamefont {J.}~\bibnamefont
  {Berges}}, \bibinfo {author} {\bibfnamefont {A.}~\bibnamefont {Rothkopf}}, \
  and\ \bibinfo {author} {\bibfnamefont {J.}~\bibnamefont {Schmidt}},\ }\href
  {\doibase 10.1103/PhysRevLett.101.041603} {\bibfield  {journal} {\bibinfo
  {journal} {Phys. Rev. Lett.}\ }\textbf {\bibinfo {volume} {101}},\ \bibinfo
  {pages} {041603} (\bibinfo {year} {2008})}\BibitemShut {NoStop}%
\bibitem [{\citenamefont {Trotzky}\ \emph {et~al.}(2008)\citenamefont
  {Trotzky}, \citenamefont {Cheinet}, \citenamefont {F{\"o}lling},
  \citenamefont {Feld}, \citenamefont {Schnorrberger}, \citenamefont {Rey},
  \citenamefont {Polkovnikov}, \citenamefont {Demler}, \citenamefont {Lukin},\
  and\ \citenamefont {Bloch}}]{2008amosuperexchange}%
  \BibitemOpen
  \bibfield  {author} {\bibinfo {author} {\bibfnamefont {S.}~\bibnamefont
  {Trotzky}}, \bibinfo {author} {\bibfnamefont {P.}~\bibnamefont {Cheinet}},
  \bibinfo {author} {\bibfnamefont {S.}~\bibnamefont {F{\"o}lling}}, \bibinfo
  {author} {\bibfnamefont {M.}~\bibnamefont {Feld}}, \bibinfo {author}
  {\bibfnamefont {U.}~\bibnamefont {Schnorrberger}}, \bibinfo {author}
  {\bibfnamefont {A.~M.}\ \bibnamefont {Rey}}, \bibinfo {author} {\bibfnamefont
  {A.}~\bibnamefont {Polkovnikov}}, \bibinfo {author} {\bibfnamefont {E.~A.}\
  \bibnamefont {Demler}}, \bibinfo {author} {\bibfnamefont {M.~D.}\
  \bibnamefont {Lukin}}, \ and\ \bibinfo {author} {\bibfnamefont
  {I.}~\bibnamefont {Bloch}},\ }\href {\doibase 10.1126/science.1150841}
  {\bibfield  {journal} {\bibinfo  {journal} {Science}\ }\textbf {\bibinfo
  {volume} {319}},\ \bibinfo {pages} {295} (\bibinfo {year}
  {2008})}\BibitemShut {NoStop}%
\bibitem [{\citenamefont {Yan}\ \emph {et~al.}(2013)\citenamefont {Yan},
  \citenamefont {Moses}, \citenamefont {Gadway}, \citenamefont {Covey},
  \citenamefont {Hazzard}, \citenamefont {Rey}, \citenamefont {Jin},\ and\
  \citenamefont {Ye}}]{2013amodipolarexchange}%
  \BibitemOpen
  \bibfield  {author} {\bibinfo {author} {\bibfnamefont {B.}~\bibnamefont
  {Yan}}, \bibinfo {author} {\bibfnamefont {S.~A.}\ \bibnamefont {Moses}},
  \bibinfo {author} {\bibfnamefont {B.}~\bibnamefont {Gadway}}, \bibinfo
  {author} {\bibfnamefont {J.~P.}\ \bibnamefont {Covey}}, \bibinfo {author}
  {\bibfnamefont {K.~R.~A.}\ \bibnamefont {Hazzard}}, \bibinfo {author}
  {\bibfnamefont {A.~M.}\ \bibnamefont {Rey}}, \bibinfo {author} {\bibfnamefont
  {D.~S.}\ \bibnamefont {Jin}}, \ and\ \bibinfo {author} {\bibfnamefont
  {J.}~\bibnamefont {Ye}},\ }\href {https://doi.org/10.1038/nature12483}
  {\bibfield  {journal} {\bibinfo  {journal} {Nature}\ }\textbf {\bibinfo
  {volume} {501}},\ \bibinfo {pages} {521} (\bibinfo {year}
  {2013})}\BibitemShut {NoStop}%
\bibitem [{\citenamefont {Parker}\ \emph {et~al.}(2013)\citenamefont {Parker},
  \citenamefont {Ha},\ and\ \citenamefont {Chin}}]{2013amoferromagnet}%
  \BibitemOpen
  \bibfield  {author} {\bibinfo {author} {\bibfnamefont {C.~V.}\ \bibnamefont
  {Parker}}, \bibinfo {author} {\bibfnamefont {L.-C.}\ \bibnamefont {Ha}}, \
  and\ \bibinfo {author} {\bibfnamefont {C.}~\bibnamefont {Chin}},\ }\href
  {https://doi.org/10.1038/nphys2789} {\bibfield  {journal} {\bibinfo
  {journal} {Nature Physics}\ }\textbf {\bibinfo {volume} {9}},\ \bibinfo
  {pages} {769} (\bibinfo {year} {2013})}\BibitemShut {NoStop}%
\bibitem [{\citenamefont {Hung}\ \emph {et~al.}(2016)\citenamefont {Hung},
  \citenamefont {Gonz{\'a}lez-Tudela}, \citenamefont {Cirac},\ and\
  \citenamefont {Kimble}}]{2016hungexchange}%
  \BibitemOpen
  \bibfield  {author} {\bibinfo {author} {\bibfnamefont {C.-L.}\ \bibnamefont
  {Hung}}, \bibinfo {author} {\bibfnamefont {A.}~\bibnamefont
  {Gonz{\'a}lez-Tudela}}, \bibinfo {author} {\bibfnamefont {J.~I.}\
  \bibnamefont {Cirac}}, \ and\ \bibinfo {author} {\bibfnamefont {H.~J.}\
  \bibnamefont {Kimble}},\ }\href {\doibase 10.1073/pnas.1603777113} {\bibfield
   {journal} {\bibinfo  {journal} {Proceedings of the National Academy of
  Sciences}\ }\textbf {\bibinfo {volume} {113}},\ \bibinfo {pages} {E4946}
  (\bibinfo {year} {2016})},\ \Eprint
  {http://arxiv.org/abs/https://www.pnas.org/content/113/34/E4946.full.pdf}
  {https://www.pnas.org/content/113/34/E4946.full.pdf} \BibitemShut {NoStop}%
\bibitem [{\citenamefont {Davis}\ \emph {et~al.}(2019)\citenamefont {Davis},
  \citenamefont {Bentsen}, \citenamefont {Homeier}, \citenamefont {Li},\ and\
  \citenamefont {Schleier-Smith}}]{2019amospin}%
  \BibitemOpen
  \bibfield  {author} {\bibinfo {author} {\bibfnamefont {E.~J.}\ \bibnamefont
  {Davis}}, \bibinfo {author} {\bibfnamefont {G.}~\bibnamefont {Bentsen}},
  \bibinfo {author} {\bibfnamefont {L.}~\bibnamefont {Homeier}}, \bibinfo
  {author} {\bibfnamefont {T.}~\bibnamefont {Li}}, \ and\ \bibinfo {author}
  {\bibfnamefont {M.~H.}\ \bibnamefont {Schleier-Smith}},\ }\href {\doibase
  10.1103/PhysRevLett.122.010405} {\bibfield  {journal} {\bibinfo  {journal}
  {Phys. Rev. Lett.}\ }\textbf {\bibinfo {volume} {122}},\ \bibinfo {pages}
  {010405} (\bibinfo {year} {2019})}\BibitemShut {NoStop}%
\bibitem [{\citenamefont {Hauschild}\ and\ \citenamefont
  {Pollmann}(2018)}]{2018tenpy}%
  \BibitemOpen
  \bibfield  {author} {\bibinfo {author} {\bibfnamefont {J.}~\bibnamefont
  {Hauschild}}\ and\ \bibinfo {author} {\bibfnamefont {F.}~\bibnamefont
  {Pollmann}},\ }\href {\doibase 10.21468/SciPostPhysLectNotes.5} {\bibfield
  {journal} {\bibinfo  {journal} {SciPost Phys. Lect. Notes}\ ,\ \bibinfo
  {pages} {5}} (\bibinfo {year} {2018})},\ \bibinfo {note} {code available from
  \url{https://github.com/tenpy/tenpy}},\ \Eprint
  {http://arxiv.org/abs/1805.00055} {arXiv:1805.00055} \BibitemShut {NoStop}%
\end{thebibliography}


\clearpage

\renewcommand{\thefigure}{S\arabic{figure}}
\renewcommand{\theequation}{S\arabic{equation}}
\renewcommand{\thesection}{S\arabic{section}}
\setcounter{page}{1}
\setcounter{equation}{0}
\setcounter{figure}{0}
\setcounter{section}{0}

\begin{widetext}

\begin{center}
{\large\bf Supplement for `Transverse instability and universal decay of spin spiral order in the Heisenberg model'}

\vspace{4mm}

Joaquin F. Rodriguez-Nieva$^{1}$, Alexander Schuckert$^{2,3}$, Dries Sels$^{4,5,6}$, Michael Knap$^{2,3}$, Eugene Demler$^6$

{\small\it $^1$Department of Physics, Stanford University, Stanford, CA 94306, USA}

{\small\it $^2$Department of Physics and Institute for Advanced Study,Technical University of Munich, 85748 Garching, Germany}

{\small\it $^3$Munich Center for Quantum Science and Technology (MCQST), Schellingstr. 4, D-80799 M{\"u}nchen}

{\small\it $^4$Department of Physics, New York University, New York, NY, USA}

{\small\it $^5$Center for Computational Quantum Physics, Flatiron Institute, New York, NY, USA}

{\small\it$^6$Department of Physics, Harvard University, Cambridge, MA 02138, USA}

\end{center}


\vspace{6mm}

The outline of the Supplement is as follows. In Section\,S1, we present the derivation of the Bogoliubov analysis which lead to Eq.(\ref{eq:bogoliubov}) of the main text. In Section\,S2, we analyze the limit $\bm q\rightarrow 0$ and $\theta \rightarrow 0$ of the Bogoliubov analysis and compare the results with a weakly coupled interacting Bose gas. In Section\,S3, we discuss additional numerical results that complement those in Fig.\ref{fig:relaxation}(d) of the main text. In Section\,S4, we discuss the dynamic instability in the one-dimensional Heisenberg model for $S=1/2$. 

\section{S1. Details of the Bogoliubov analysis}

We begin our analysis by parametrizing the spin degrees of freedom on the upper hemisphere of the Bloch sphere, $S_i^z = \sqrt{S^2-|S_i|^2}$. The resulting equations of motion are 
\be
\dot{S}_i^\pm = \pm i \sum_{j\in{\cal N}_i}\left[S_i^\pm\sqrt{S^2-|S_j|^2} -S_j^\pm \sqrt{S^2-|S_i|^2}\right], 
\label{seq:eom}
\ee
with energy and inverse time in units of $JS$. Taking small deviation over the steady-state solution $\bar{S}_i$, $S_i^\pm = \bar{S}_i^\pm + \delta S_i^\pm$, Eq.\,(\ref{seq:eom}) reads
\be
i\delta\dot{S}_i^\pm = \pm \sum_{j\in{\cal N}_i} \cos\theta\left[\,\delta S_j^\pm + \frac{\bar{S}_j^\pm\left(\bar{S}_i^+\delta S_i^- + \bar{S}_i^-\delta Sa_i^+\right)}{2\cos^2\theta}- (i\leftrightarrow j)\right].
\ee
Because $\bar{S_i}^\pm$ is both time and position-dependent, it is convenient to write fluctuations relative to the wavevector and frequency of $\bar{S}_i$, {\it i.e.}, $\delta S_i^\pm = e^{\pm i({\bm q}\cdot {\bm r}_i+\mu t)}\sum_{\bm k}e^{i({\bm k}\cdot {\bm r}_i+\omega_{\bm k}t)}\delta S_{\bm q \pm \bm k}^\pm$. The linearized equations of motion can thus be written as 
\be
(\omega_{\bm k}\pm\mu)\delta S_{\bm q \pm \bm k}^\pm = \mp \cos\theta\left[ \left(\gamma_0-\gamma_{\bm q \pm \bm k} - \frac{\tan^2\theta(\gamma_{\bm k} - \gamma_{\bm q})}{2}\right)\delta S_{\bm q \pm \bm k}^\pm - \frac{\tan^2\theta(\gamma_{\bm k} - \gamma_{\bm q})}{2}\delta S_{\bm q \mp \bm k}^\mp\right],
\label{eq:linearized}
\ee
with $\gamma_{\bm q}=\sum_{\bm a}e^{i{\bm q}\cdot{\bm a}}$ ($\bm a$: unit vectors of the lattice). Defining ${\varepsilon}_{\bm p}= \cos\theta(\gamma_0-\gamma_{\bm p})$ and $\Delta_{\bm k} = -\sin\theta\tan\theta(\gamma_{\bm k}-\gamma_{\bm q})$ leads to Eq.(\ref{eq:bogoliubov}) of the main text. 

Adding an anisotropic term to the Heisenberg Hamiltonian, $\delta\hat{H}=-\epsilon\sum_{\langle i j \rangle}\hat{S}_i^z\hat{S}_j^z$, and repeating the same procedure above leads to the same form of Eq.(\ref{eq:bogoliubov}) with modified energy and pairing 
\be
   {\varepsilon}_{\bm p} = \cos\theta\left[(1+\epsilon) \gamma_0 - \gamma_{\bm p}\right],    \quad{\Delta}_{\bm k} = -\sin\theta\tan\theta {\left[(1+\epsilon)\gamma_{\bm k}-\gamma_{\bm q}\right]}.
\ee
Note that, in the long-wavelength, limit, the pairing becomes hard-core and repulsive if $\epsilon<0$, {\it i.e.}, $\Delta_{\bm k} \approx |\epsilon|\sin\theta\tan\theta$. 

\subsection{Bogoliubov analysis in spherical coordinates}

The previous derivation has the disadvantage that the parametrization is singular at $\theta = \pi/2$. In order to avoid the singular behaviour in the physically interesting case  $\theta = \pi/2$, we can use polar coordinates, ${\bm\Omega}_i=(\sin\theta_i\cos\phi_i,\sin\theta_i\sin\phi_i,\cos\theta_i)$, to parametrize the spin orientation. As we will see, this does not change the end result when we take the limit $\theta \rightarrow \pi/2$. 
The equation of motion are 
\be
i \dot{\Omega}_i^\pm = \pm\sum_{j\in{\cal N}_i} [\Omega_i^\pm\Omega_j^z-\Omega_j^\pm\Omega_i^z], 
\label{eq:omegaeom}
\ee
with mean field solution 
\be
\bar{\Omega}_i^\pm = \sin\theta e^{\pm i({\bm q}\cdot{\bm r}_i-\mu t)}, \quad\bar{\Omega}_i^z = \cos\theta,  
\ee
and $\mu = \cos\theta(\gamma_0-\gamma_{\bm q})$. Fluctuations on top of the mean-field equations are parametrized by 
\be
\delta\Omega_i^\pm = e^{\pm i({\bm q}\cdot{\bm r}_i-\mu t)}\left(\pm i \delta\phi_i + \cos\theta_i \delta\theta_i\right),\quad \delta\Omega_i^z = - \sin\theta \delta\theta_i. 
\label{eq:domega}
\ee
By replacing Eq.(\ref{eq:domega}) into Eq.(\ref{eq:omegaeom}), we find 
\be
\pm\mu\cos\theta\delta\theta_i\mp\delta\dot{\phi}_i +i(\mu\delta\phi_i + \cos\theta\delta\dot{\theta}_i) = \pm\sum_{j\in{\cal N}_i}\left[ -\sin^2\theta(\delta\theta_j-e^{\pm i{\bm q}\cdot{\bm r}_{ji}}\delta\theta_i)\pm i\cos\theta(\delta\phi_i-e^{\pm i{\bm q}\cdot{\bm r}_{ji}})+\cos^2\theta(\delta\theta_i-e^{\pm i{\bm q}\cdot{\bm r}_{ji}}\delta\theta_j)\right],
\ee
where ${\bm r}_{ji} = {\bm r}_j - {\bm r}_i$ and ${\cal N}_i$ denotes the nearesest neighbor of $i$. The real and imaginary parts of this equation are given by 
\be
\begin{array}{l}
\displaystyle  \delta\dot{\phi}_i = \sum_{j\in{\cal N}_i} \left[ \sin^2\theta(\delta\theta_j-\cos({\bm q}\cdot{\bm r}_{ji})\delta\theta_i) - \cos^2\theta(\delta\theta_i-\cos({\bm q}\cdot{\bm r}_{ji})\delta\theta_j)-\cos\theta\sin({\bm q}\cdot{\bm r}_{ji})\delta\phi_j+\mu\cos\theta\delta\theta_i\right], \\ \\
\displaystyle  \delta\dot{\theta}_i = \sum_{j\in{\cal N}_i}\left[ (\delta\phi_i - \cos({\bm q}\cdot{\bm r}_{ji})\delta\phi_j)-\frac{\mu}{\cos\theta}\delta\phi_i-\cos\theta\sin({\bm q}\cdot{\bm r}_{ji})\delta\theta_j\right].
\end{array}
\ee
Going into Fourier space and using the relations $\sum_{{\bm a}}\cos({\bm q}\cdot{\bm a})e^{i{\bm k}\cdot{\bm a}} = \frac{\gamma_{\bm q+\bm k}+\gamma_{\bm q-\bm k}}{2}$ and $\sum_{{\bm a}}\sin({\bm q}\cdot{\bm a})e^{i{\bm k}\cdot{\bm a}} = \frac{\gamma_{\bm q+\bm k}-\gamma_{\bm q-\bm k}}{2i}$, results in
\be
\begin{array}{l}
\displaystyle  \delta\dot{\phi}_{\bm k} = \left[ \cos^2\theta\left(\frac{\gamma_{\bm q+\bm k}+\gamma_{\bm q-\bm k}}{2} - \gamma_{\bm q}\right)+\sin^2\theta(\gamma_{\bm k}-\gamma_{\bm q}) \right]\delta\theta_{\bm k} - \cos\theta\left(\frac{\gamma_{\bm q+\bm k}-\gamma_{\bm q - \bm k}}{2i}\right)\delta\phi_{\bm k}, \\
\displaystyle \delta\dot{\theta}_{\bm k} = -\left(\frac{\gamma_{\bm q+\bm k}+\gamma_{\bm q-\bm k}}{2}-\gamma_{\bm q}\right)\delta\phi_{\bm k} - \cos\theta\left(\frac{\gamma_{\bm q + \bm k} - \gamma_{\bm q-\bm k}}{2i}\right)\delta\theta_{\bm k}. 
\end{array}
\ee
If we define $\Delta\varepsilon' = \gamma_{\bm q+\bm k}+\gamma_{\bm q-\bm k}-2\gamma_{\bm q}$ and $\Delta_{\bm k}' = \gamma_{\bm k} - \gamma_{\bm q}$. The imaginary part of the eigenvalue equation is given by 

\be
{\rm Im}[\omega_{\bm k}] = \frac{1}{2}\sqrt{\Delta\varepsilon'[\Delta\varepsilon'\cos^2\theta+\sin^2\theta \Delta_{\bm k}']}
\ee
which coincides with the expression of the main text after identifying $\Delta\varepsilon = \Delta\varepsilon' \cos\theta$ and $\Delta_{\bm k}=-\sin\theta\cos\theta\Delta_{\bm k}'$.

\section{S2. Bogoliubov analysis on the long-wavelength theory}

To make the connection with the usual BEC theory more crisp, we make a one-to-one comparison between the long wavelength effective theory of the Heisenberg model and a weakly interacting Bose gas in the limit $\bm q \rightarrow 0$ and $\theta \rightarrow 0$. This analysis also shows that the results above are a generic feature of SU(2) symmetry rather than a peculiarity of the nearest-neighbour Heisenberg model in Eq.(\ref{eq:hamiltonian}), and that the underlying lattice is not essential as in the modulational instability. Assuming small deviations from the ferromagnetic ground state $|F\rangle$ and performing a Holstein-Primakoff transformation $\hat{S}_i^+ = \sqrt{2S-\hat{\psi}_{i}^\dagger\hat{\psi}_{i}}\hat{\psi}_i$ and $\hat{S}_i^z = S -\hat{\psi}_i^\dagger\hat{\psi}_i$, to quartic order in the bosonic operators $\hat{\psi}_i$ leads to the long-wavelength Hamiltonian
\be
\hat{H} = JSa^2 \int_{\bm x} \left(\nabla\hat{\psi}_{\bm x}^\dagger \nabla\hat{\psi}_{\bm x} + \frac{1}{4S}\hat{\psi}_{\bm x}^\dagger\hat{\psi}_{\bm x}^\dagger\nabla\hat{\psi}_{\bm x}\nabla\hat{\psi}_{\bm x}+{\it h.c.}\right).
\label{eq:Heffective}
\ee
Unlike the usual Bose gas with hard core collisions, here the collision amplitude of two quasiparticles with momentum ${\bm k}$ and ${\bm p}$ is $\propto - ({\bm k}\cdot{\bm p})$. This reflects the SU(2) symmetry of the Hamiltonian: collisions become negligible at small momenta because a ${\bm k}\rightarrow 0$ magnon state, $\hat{\psi}_{\bm k}^\dagger|{\rm F}\rangle \approx \frac{\hat{S}_{\bm k}^+}{\sqrt{2S}}|{\rm F}\rangle$, is effectively a global rotation of $|{\rm F}\rangle$ that would not affect the dynamics of a second incoming magnon. Furthermore, unlike the BEC theory that contains a characteristic velocity $v_* = \sqrt{gn/m}$ that quantifies the sound velocity of linearly-dispersing quasiparticles and the resulting Landau criterion ($n$: condensate density, $m$: mass), there is no emergent velocity in Eq.(\ref{eq:Heffective})---this results in the well-known fact that the Goldstone modes of the ferromagnet do not have sound-like dispersion.

In the presence of a condensate $\langle\hat{\psi}_{\bm x}^\pm\rangle = \psi_0e^{\pm i (\bm q\cdot {\bm x} + \mu t)}$, linearization of the equation of motion leads to 
\be
i\partial_t\delta{\boldsymbol\psi} = \left(\begin{array}{cc} \varepsilon_{\bm q + \bm k}^0-2Ja^2{\bm q}\cdot(\bm q+\bm k)|\psi_0|^2-\mu & -Ja^2\psi_0^2(\bm q + \bm k)\cdot(\bm q - \bm k) \\ Ja^2\psi_0^2(\bm q + \bm k)\cdot(\bm q - \bm k) & -\varepsilon_{\bm q - \bm k}^0 + 2Ja^2\psi_0^2{\bm q}\cdot(\bm q - \bm k)\end{array}\right) \delta{\boldsymbol\psi}, 
\label{eq:bogoliuboveffective}
\ee
where $\delta{\boldsymbol\psi}^t = (\delta\psi_{\bm q+\bm k},\delta\bar{\psi}_{\bm k - \bm q})$, $\mu$ is the chemical potential $\mu = JSa^2{\bm q}^2 - Ja^2\psi_0^2{\bm q}^2$, and $\varepsilon_{\bm p}^0$  is the bare magnon energy defined as $\varepsilon_{\bm p}^0 = JSa^2{\bm p}^2$. First, we re-arrange the terms in the diagonal of the matrix: 
\be
\varepsilon_{\bm q\pm \bm k}^0 -2Ja^2\psi_0^2{\bm q}\cdot(\bm q \pm\bm k)-\mu=\varepsilon_{\bm q\pm \bm k}^0 -Ja^2\psi_0^2(\bm q \pm\bm k)^2-Ja^2\psi_0^2({\bm q}^2-{\bm k}^2)-\mu.   
\ee
Second, we define the renormalized energy $\varepsilon_{\bm q\pm\bm k} = \varepsilon_{\bm q\pm \bm k}^0-Ja^2\psi_0^2({\bm q\pm \bm k})^2$ and the binding energy $\Delta_{\bm k}/2 = -Ja^2\psi_0^2({\bm q}^2-{\bm k}^2)$. The value of $\psi_0$ is related to $\theta$ through $\sqrt{2S-\psi_0^2}\psi_0 = S\sin\theta$, or $\psi_0 \approx\sqrt{\frac{S}{2}}\sin\theta$. This results in the pairing energy $\Delta_{\bm k} = - JSa^2\sin^2\theta({\bm q}^2 - {\bm k}^2)$, consistent with the long-wavelength expansion of Eq.(\ref{eq:parameters}) of the main text.

Terms that reduce the symmetry from SU(2) to U(1), such as the exchange anisotropy $\epsilon<0$ in Eq.(\ref{eq:hamiltonian}), give rise to hard-core collisions in Eq.(\ref{eq:Heffective}) with strength $g = -\epsilon J$, leading to a repulsive and hard-core pairing $\Delta_{\bm k} = |\varepsilon|J\sin^2\theta$ in the easy-plane case.

\begin{figure}[t]
  \includegraphics[scale = 1.0]{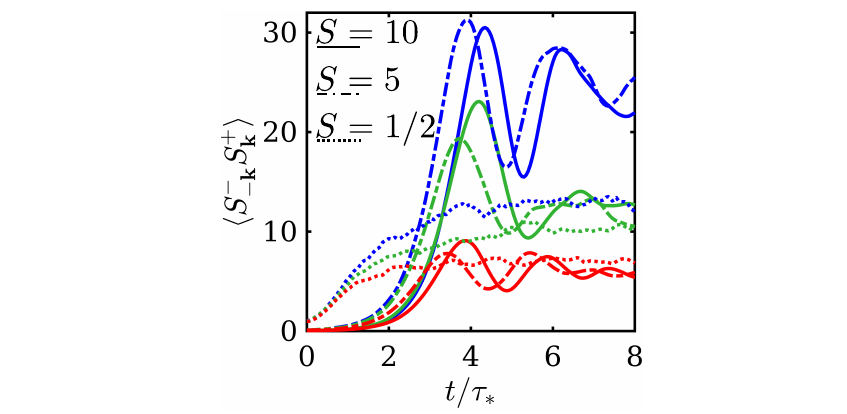}
  \vspace{-3mm}
  \caption{Growth of the most unstable mode showing saturation and subsequent oscillations plotted for different initial conditions and spin number $S$. The colorcode indicates simulations with different $\bm q$: ${q}_xa = 0.25$ (blue), $q_xa = 0.37$ (green), and $q_xa = 0.5$ (red). The linestyle indicates models with different value of $S$. All figures use $\theta = \pi/2$. } 
\label{fig:supp_growth}
\end{figure}

\section{S3. Saturation of the dynamic instability}

In the main text we argued that the instability growth is self-regulated by the spin number constraint, and that fluctuations peak at $t \approx 4\tau_*$ irrespective of the initial state, see definition of $\tau_*$ in Eq.(\ref{eq:scaling}). In Figure \ref{fig:supp_growth}, we show this behavior for various initial conditions and values of the spin number $S$. 

\section{S4. Transverse instability for spin 1/2 in 1D}

Here we show that the dynamics described in the main text for two-dimensional spin spirals in the large $S$ limit are also relevant for $S=1/2$ in 1D. We focus on full spirals with $\theta = \pi/2$. In Fig.~\ref{fig:both} we display the connected spin correlation functions
\be
\begin{array}{rl}
C^{zz}_k(t) & \equiv \langle{\hat S^z_k (t) \hat S^z_{-k} (t)}\rangle,\\
C^{+-}_k(t) & \equiv \langle{\hat S^+_k (t) \hat S^-_{-k} (t)}\rangle-\langle{\hat S^+_k (t)}\rangle\langle{\hat S^-_{-k} (t)}\rangle,
\end{array}
\label{eq:CzzCpm}
\ee
for a spiral wavelength $Q = 0.13\pi$ obtained from MPS-iTEBD simulations employing the TeNPy package~\cite{2018tenpy}, with a unit cell chosen large enough to fit the spiral (here: $L=90$). For very short times, pertubartive short time dynamics dominate, see derivation below. At around $Jt=5$, the dynamic instability takes over, leading to a growth of fluctuations with momenta around the spiral wavelength, $k=\pm Q$.  The distribution of fluctuations in momentum space approximately agree with the imaginary part of the Bogoliobov dispersion~\cite{2014spiralexp}. We find similar signatures of the instability in the $C^{+-}$ correlations. However, we find the that low-momentum part ($k\le Q$) of the ``double lobe'' structure of the Bogoliubov dispersion dominates over the large-momentum part ($k>Q$).

\begin{figure}[ht]
\includegraphics[scale=0.8]{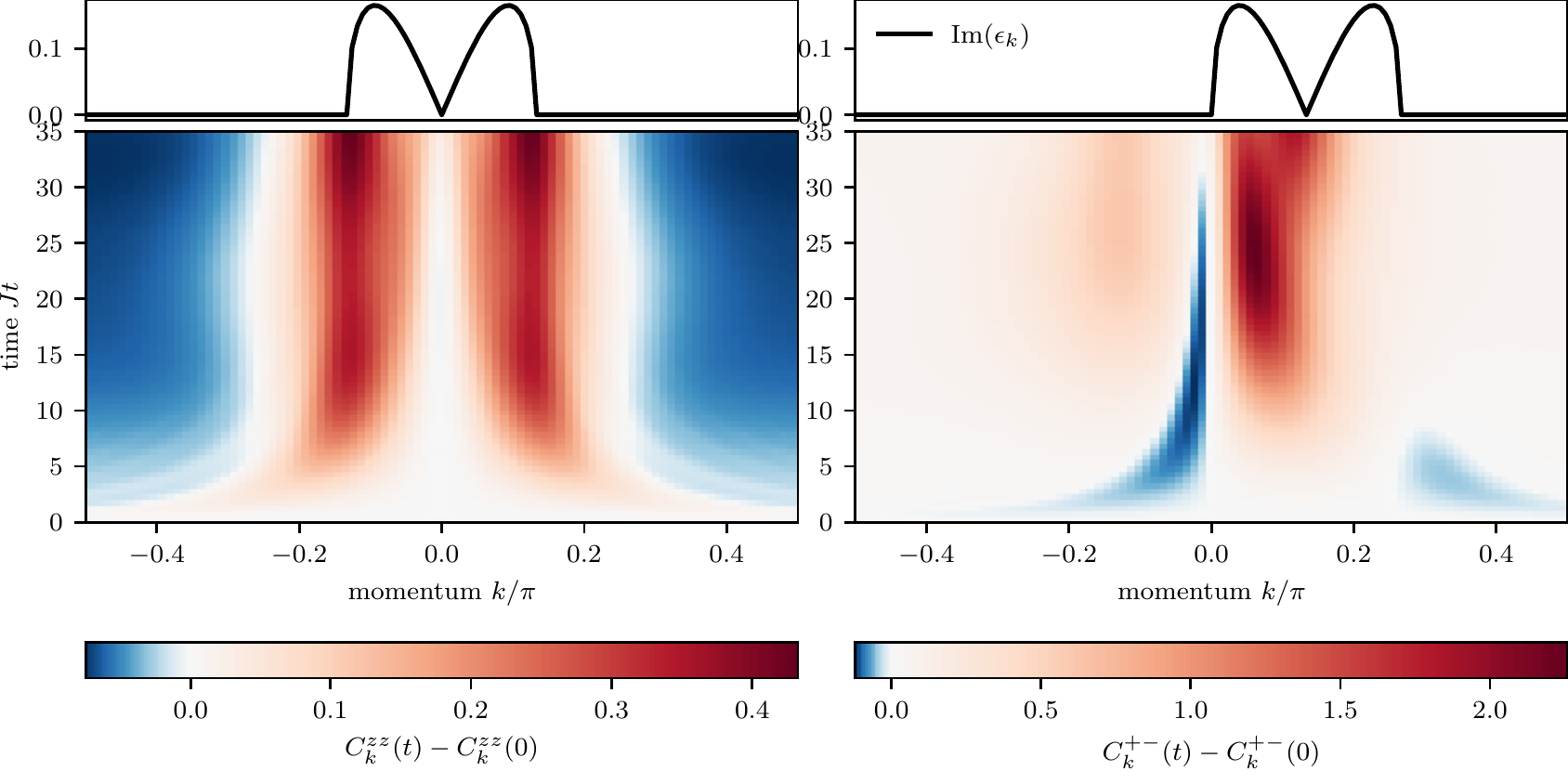}
\caption{\label{fig:both} {Growth of spiral fluctations in 1D.} Shown are the $C_{\bm k}^{zz}$ and $C_{\bm k}^{+-}$ correlations for a spiral wavelength $Q = 0.13\pi$, obtained from iTEBD-MPS simulations. The upper panels show the results of the Bogoliubov analysis.} 
\end{figure}

\begin{figure}[ht]
\includegraphics[scale=1]{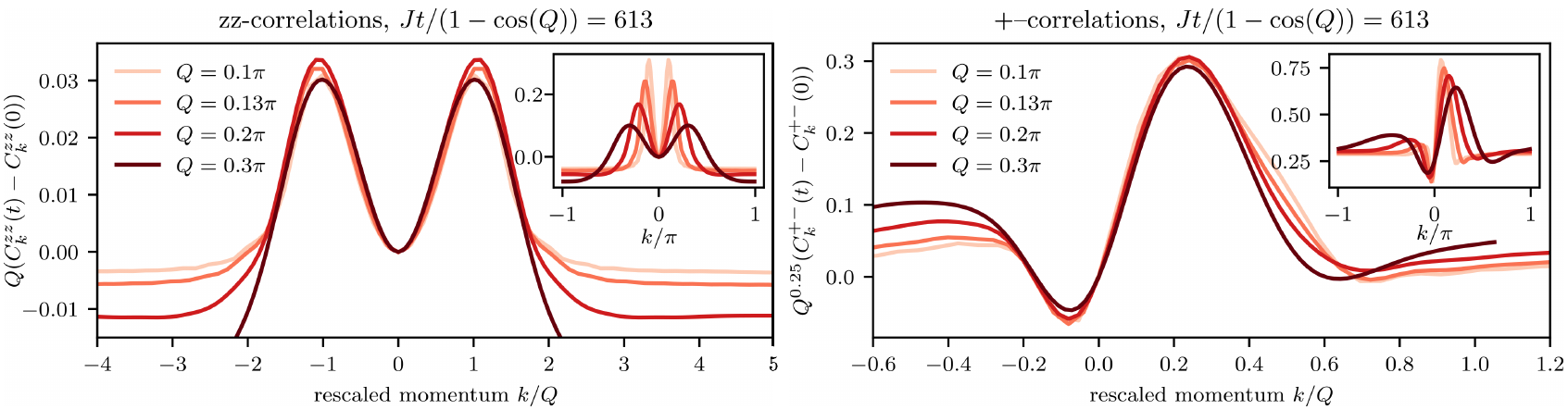}
\caption{\label{fig:univ}   Spatial-temporal scaling of spin fluctuations. Shown is the $C^{+-}$ and $C^{zz}$ correlations as defined in Eq.~\eqref{eq:CzzCpm} rescaled according to the scaling function in Eq.~\eqref{eq:scalingfct}, with the rescaled time $Jt/(1-\cos(Q))$ fixed.  Insets: Unrescaled correlation functions.} 
\end{figure}

In analogy with the scaling of fluctuations shown in in Fig.\ref{fig:relaxation}(c) of the main text, we also find scaling in the growth of fluctuations. As shown in Fig.~\ref{fig:univ}, the scaling relation is given by
\be
C^{ab}_k\left(\frac{Jt}{1-\cos(Q)}\right)-C^{ab}_k(0)=Q^{-\alpha} f\left(\frac{k}{Q^{\beta}}\right).
\label{eq:scalingfct}
\ee
While the time and momentum rescaling factors $\beta=1$ are obtained analytically from the Bogoliubov treatment, we find $\alpha\approx 0.25$ for $ab=+-$ and $\alpha\approx 1$ for $ab=zz$ numerically from our MPS simulations. Note that our results can not be explained by a simple perturbative short time scaling. Indeed, we analytically show next that the perturbative short time scale is \emph{Q-independent}, in contrast to the $k/Q$ dependence we find here.

To understand the short time behavior, we consider the general Heisenberg model
\be
\hat{H}= \sum_{i\neq j} J_{ij}\left(\hat{S}^+_i\hat{S}^-_{j}+\hat{S}^z_i \hat{S}^z_{j}\right),
\ee
where $\hat{S}^+_i=\hat{S}^x_i+i \hat{S}^y_i$ and we assume real couplings $J_{ij}=J_{ji}$. We calculate the short time dynamics of the zz-fluctuations $C_{\bm k}^{zz}=\frac{1}{N}\sum_{i,j} e^{i{\bm k}({\bm r}_i-{\bm r}_j)} \langle{\Psi(t)| \hat{S}^z_i \hat{S}^z_j|\Psi (t)}\rangle$ by expanding 
\be
C_{\bm k}^{zz}(t) \approx C_{\bm k}^{zz}(0) - \frac{t^2}{2}\frac{1}{N}\sum_{i,j} e^{i{\bm k}({\bm r}_i-{\bm r}_j)} \langle{\Psi(0)|[\hat{H},[\hat{H}, \hat{S}^z_i \hat{S}^z_j]]|\Psi(0)}\rangle.
\ee

The expression can be evaluated easiest for the spiral initial state by introducing the unwinding transformation $U$ as $\langle{\Psi(0)|U^\dagger U[\hat{H},[\hat{H}, \hat{S}^z_i \hat{S}^z_j]]U^\dagger U|\Psi(0)}\rangle$ with $U|{\Psi(0)}\rangle=|{\uparrow\cdots\uparrow}_x\rangle$ and $US^{\pm}_iU^\dagger=e^{\mp i{ Q}{ r}_i}S^{\pm}_i$, $US^{z}_iU^\dagger=S^{z}_i$ \emph{after} calculating the double commutator. The general result is
\be
\begin{array}{ll}
\displaystyle \langle{\Psi(0)|[\hat H,[\hat H, \hat S^z_k \hat S^z_l]]|\Psi(0)}\rangle= & \displaystyle -\frac{1}{4}J_{kl}\sum_j(J_{kj}+J_{jl})(\cos({Q}{r}_{jk})+\cos({Q}{r}_{jl})\notag \\ \\
& \displaystyle \quad+\frac{1}{2}J_{kl}^2\left(\cos({Q}{r}_{kl})-\cos^2({Q}{r}_{kl})\right)\notag\\ \\
&\displaystyle \quad+\frac{1}{4}\sum_j J_{lj}J_{jk}\left(\cos({Q}{r}_{jl})+\cos({Q}{r}_{jk})-2\cos({Q}{r}_{jk})\cos({Q}{r}_{jl})\right)\notag\\ \\
&\displaystyle \quad+\frac{1}{2}J_{kl}\cos({Q}{r}_{kl})\left(\sum_jJ_{lj} \cos({Q}{r}_{jl})+\sum_jJ_{jk}\cos({Q}{r}_{jk})\right),
\end{array}
\ee
where ${r}_{jk}={r}_j-{r}_{k}$. Evaluating this expression for 1D nearest neighbour interactions $J_{ij}=\delta_{j,i+1}+\delta_{j,i-1}$, we find
 \begin{align}
C_{k}^{zz}(t)\approx \frac{1}{4} &- \frac{t^2}{2}\bigg\lbrace \left(\cos(k)-\cos(2k)\right) \left(\cos^2 Q-\cos Q \right)\bigg\rbrace.
\label{eq:ste1d}
\end{align} 
The short time scale in angle brackets vanishes for both $k=0$ and $Q=0$ as expected and agrees with an exact MPS simulation at short times, see Fig.~\ref{fig:comp}. Importantly, the location of the maximum of the short time scale is \emph{independent} of $Q$ and located at $k\approx \pm 0.42\pi$.
\begin{figure}
\includegraphics[scale=0.8]{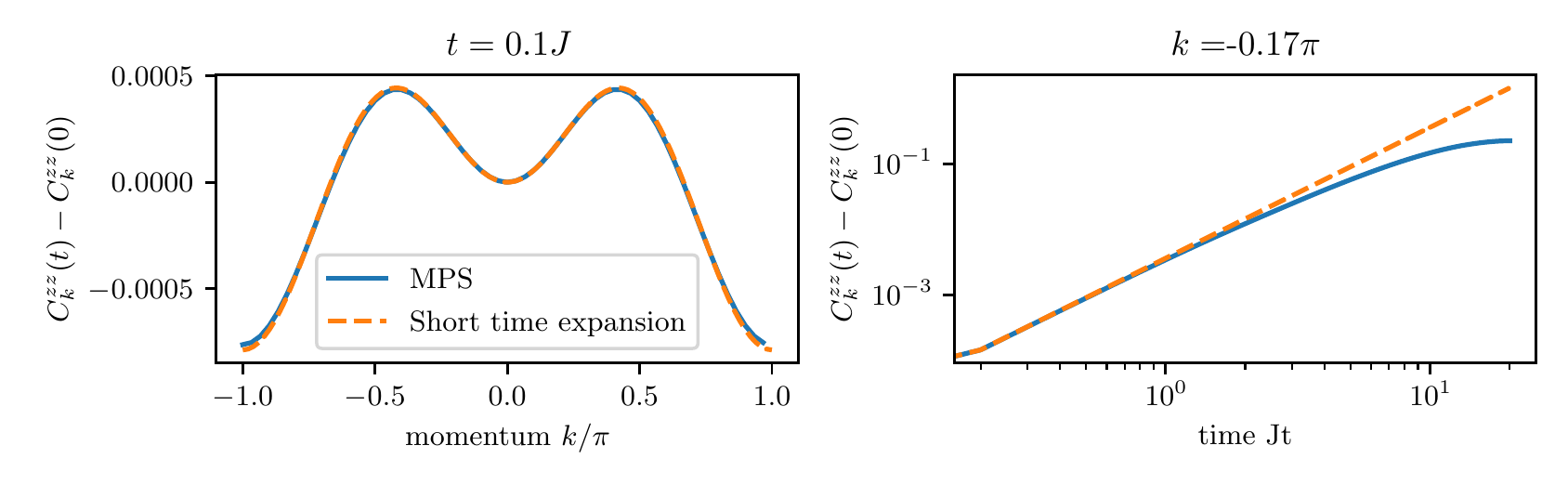}
\caption{\label{fig:comp} Comparison between Eq.~(\ref{eq:ste1d}) and an iTEBD simulation for $Q=0.13\pi$.} 
\end{figure}

\end{widetext}

\end{document}